# SF$_6$ streamer breakdown induced by floating linear metal particles: Following streamers and side streamers

Zihao Feng©,* Liyang Zhang, Xinxin Wang, Xiaobing Zou©, Haiyun Luo, and Yangyang Fu©†

*Department of Electrical Engineering, Tsinghua University, Beijing 100084, China*

The electrical breakdown of SF$_6$ in the presence of floating metal particles is facilitated by two key factors: the role of floating metal particles and the nonlinear breakdown behavior of high-pressure SF$_6$. However, the microscopic transient processes remain unclear, motivating this paper. Using two-dimensional fluid models, we investigate SF$_6$ streamer breakdown induced by a floating linear metal particle under negative applied voltage. First, we identify a characteristic double-end streamer inception in the combined gap. Then, we propose the following-streamer (FS) mechanism to explain the metal particle's role. Two following streamers, FS1 and FS2, arise from the interaction between space charge and metal particle. FS1 facilitates breakdown via the negative space-charge field generated by its head. FS2 facilitates breakdown by merging with FS1, accelerating its propagation and enhancing the electric field at the primary streamer head. Finally, we propose the side-streamer (SS) mechanism to explain the nonlinear breakdown behavior of high-pressure SF$_6$. The SS is identified as a new forward ionization wave that develops along the sides of the primary streamer, due to photoionization-driven negative-ion accumulation. SS facilitates breakdown by merging with the primary streamer, increasing negative space charge and leading to three distinct propagation modes. Higher pressure increases the production rate of negative ions along the streamer sides, making SS more likely to form. Under overvoltage, the facilitating effect of SS diminishes as the background field $(E/N)_b$ strengthens, disappearing when $(E/N)_b$ exceeds 245 Td. This study provides new insights into the SF$_6$ streamer breakdown mechanisms induced by floating metal particles and offers theoretical references for further investigation on the quantitative characterization.

## I. INTRODUCTION

Gas-insulated electrical equipment, such as gas-insulated switchgear (GIS) and gas-insulated transmission lines (GIL), has seen widespread use in recent years. However, SF$_6$ discharge faults caused by floating metal particles within the sealed chambers pose a significant risk to the safe operation of these systems [1,2]. This issue has emerged as a critical weakness in the insulation performance of gas-insulated equipment [3,4]. The electrical breakdown of SF$_6$ in the presence of metal particles is facilitated by two key factors: the role of floating metal particles and the nonlinear breakdown behavior of high-pressure SF$_6$. While previous research has examined these factors through experimental and analytical approaches, a deeper understanding of the transient microscopic characteristics remains, which motivates the research presented in this paper.

Regarding the role of floating metal particles, extensive experimental research has consistently shown that metal particles significantly reduce the gap breakdown voltage [5–9], with linear-shaped particles with protrusions having a greater impact compared to other shapes [10,11]. Recently, microdischarge theory [12–16] has been commonly cited to explain the role of metal particles in facilitating breakdown. This theory is based on the short-circuit effect [17], suggesting that a particle-induced microdischarge in a narrower gap acts as a protrusion on the electrode surface. However, the short-circuit assumptions do not fully capture the microscopic processes involved. Specifically, when a particle-induced discharge occurs, the interaction between the space charge and the metal particle redistributes the surface-charge distribution on the particle [18], altering the local electric field. This, in turn, influences discharge propagation and affects breakdown behavior, but the exact role of the metal particles remains unclear.

Regarding the nonlinear breakdown behavior of SF$_6$, extensive experimental research has shown that its breakdown voltage varies nonlinearly with increasing gas pressure, particularly in scenarios involving electrode protrusions [19–23] and floating linear metal particles [24–26]. Some research has proposed physical mechanisms to explain this nonlinear behavior. For instance, experimental research by Gallimberti and Wiegart [27], Seeger and Clemen [20], and Zhao et al. [28,29] identified precursors originating from the side of the $SF_6$ streamer channel that promote leader formation and facilitate breakdown. Wu et al. [30] conducted experimental observations of a positive glow corona within the $SF_6$ streamer channel, whose shielding effect induced a side spark path and facilitated breakdown. Simulations by Feng et al. [18] showed that negative-ion accumulation near protrusions, secondary streamer, and recovery of the channel field might influence $SF_6$ breakdown. However, these simulations were limited to a narrow range of gas pressures. Meng et al. [31] simulated macroscopic characteristics of $SF_6$ streamers under varying pressure conditions, but the streamer channel morphology was not discussed in their paper, which is crucial for understanding the phenomena observed in high-pressure experiments.

___
*Contact author: zihaofeng1998@163.com
†Contact author: fuyangyang@tsinghua.edu.cn

To date, a precise understanding of these two key factors, particularly their microscopic transient behaviors, remains unclear. In this paper, we use two-dimensional (2D) axisymmetric fluid models to investigate $SF_6$ streamer breakdown induced by a floating linear metal particle under negative applied voltage, with a detailed description of the model provided in Sec. II. In Sec. III, we investigate the inception of discharge in the combined gap, focusing on the characteristic double-end streamer and analyze the underlying mechanisms driving its formation. In Sec. IV, we propose the following-streamer (FS) mechanism to illustrate the exact role of floating metal particles in facilitating breakdown. Section IV A explores the dynamics of the FS and the mechanism driving its formation, while Sec. IV B examines how the FS facilitates the breakdown process. In Sec. V, we propose the side-streamer (SS) mechanism to illustrate the nonlinear behavior of high-pressure $SF_6$ and its role in facilitating breakdown. Section V A examines the dynamics of the SS, while Sec. V B reveals its formation mechanism and how it facilitates breakdown. Section V C presents a qualitative analysis of the pressure dependence of SS, providing insights into why it is more likely to occur at high pressure. Section V D investigates the dynamics of the SS under overvoltage conditions and its influence on the breakdown process.

## II. MODEL DESCRIPTION

### A. Two-dimensional axisymmetric geometry

The discharge electrodes are modeled as parallel plates separated by a gas gap distance $d = 5$ mm, as shown in Fig. 1. A linear metal particle, with length $L = 1$ mm and diameter $a = 0.2$ mm, is positioned near the high-voltage (HV) electrode. This position reflects a typical scenario where gap breakdown occurs due to floating particles [7,12,15]. The gas gap between the electrodes is divided into two regions: a short gap ($g_s = 1$ mm) between the particle's top tip and the HV electrode, and a long gap ($g_l = 3$ mm) between the particle's bottom tip and the grounded electrode. The computational domain includes a sufficiently wide horizontal size to minimize edge effects. However, it is important to note that the simplified geometric structure used in this model differs from the real-world scenario in terms of electrode configurations and dimensions. The following discussions clarify the limitations arising from these differences.

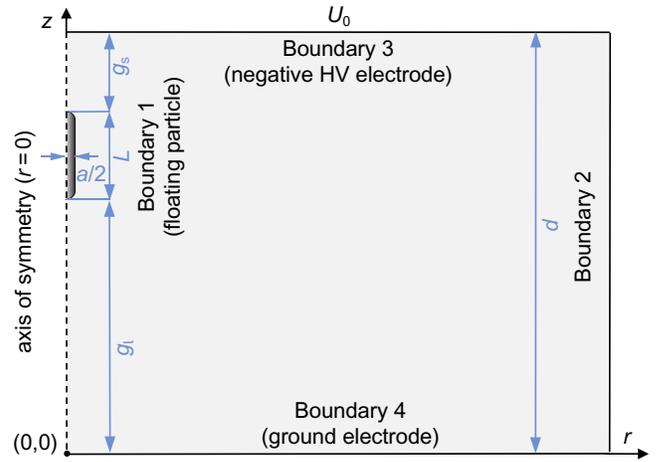

FIG. 1. Geometry of the 2D axisymmetric simulation domain and the computational boundaries.

Regarding the electrode configuration, to accurately represent the field enhancement of the particle, this paper employs an axisymmetric coordinate system. However, it cannot accurately model both the coaxial cylindrical electrode structure commonly used in GIL pipelines and the particle within the same coordinate system. As a result, the electrode configuration is further simplified to parallel plates.

Regarding dimensions, to reduce computational costs, this paper employs a reduced-sized geometry, particularly disproportionately shortening the long gap $g_l$ compared to the real-world scenario. This reduction exaggerates the relative size of the floating particle compared to the whole space. Nonetheless, this simplification meets the research objective, which is to explore the general physics during the streamer stage rather than provide precise engineering predictions.

### B. Governing equations

The $SF_6$ streamer is considered a self-consistent, continuous conductive fluid. However, it should be acknowledged that the fluid assumption may not fully capture the kinetic characteristics [32–35] or stochastic phenomena, such as branching [36–38]. The fully kinetic model [39–41] or the hybrid model [42,43] provide higher accuracy.

Nonetheless, the fluid model remains suitable for exploring the general physics of streamer phenomena and can significantly reduce computational costs [44].

In the fluid model, the number density of species is obtained by solving the continuity equation,

$$\frac{\partial n_i}{\partial t} + \nabla \cdot \mathbf{\Gamma}_i = S_i, \quad (1)$$

where $n_i$, $\mathbf{\Gamma}_i$, and $S_i$ denote the number density, number flux, and source term of the $i$th species, respectively.

The drift-diffusion approximation is used to determine $\mathbf{\Gamma}_i$. This approach relies on two assumptions [45]: (1) the inertia terms in the momentum equations for species are negligible, and (2) the characteristic time scale of momentum transfer is shorter than that of streamer propagation. Consequently, one can obtain

$$\mathbf{\Gamma}_i = n_i \mathbf{u}_i = z_i \mu_i n_i \mathbf{E} - D_i \nabla n_i, \quad (2)$$

where $\mathbf{u}_i$, $z_i$, $\mu_i$, and $D_i$ denote the flux velocity, charge, mobility, and diffusion coefficient of the $i$th species, respectively; and $\mathbf{E}$ denotes the electric field.

The Poisson's equation is used to calculate the distribution of electric potential $U$:

$$\nabla^2 U = -\frac{e_0}{\varepsilon_0} \sum_i z_i n_i, \quad (3)$$

where $e_0 > 0$ is the elementary charge, and $\varepsilon_0$ is the vacuum permittivity. The electric field is calculated based on its defining equation:

$$\mathbf{E} = -\nabla U. \quad (4)$$

The local mean energy approximation (LMEA) effectively accounts for the nonlocal relationship between electron kinetics and the reduced electric field ($E/N$) near the computational boundaries ($N$ denotes the number density of the neutral species) [45,46]. Namely, the interaction between the space charge and the floating metal particle can be ensured. Therefore, the LMEA framework is employed to investigate the role of floating metal particles in Secs. III and IV. To reduce computational costs while ensuring sufficient accuracy to meet the research objectives in Sec. V, the local field approximation (LFA) is employed to investigate the nonlinear breakdown behavior of high-pressure $SF_6$. The governing equations for these two frameworks are discussed below. The finite-element method is employed to solve the equations using an implicit solver. For better reproducibility of the results presented in the main text, the details of the grid spacing and time-step scheme are provided in Appendix A.

*1. Local mean energy approximation*

In the context of the LMEA, the inclusion of the electron energy relaxation process is essential for accurately calculating electron dynamics and plasma chemistry [47,48]. The key chemical species considered in the LMEA framework include $e$, $SF_5^+$, $SF_6^-$, $SF_5^-$, $F^-$, $SF_6$, $SF_5$, F, and $SF_6^*$. Here, $SF_6^*$ represents excited electronic and vibrational levels of $SF_6$, treated as a single species. $SF_6$ serves as the background gas, and its number density is determined using the ideal gas law:

$$P = n_{SF_6} k_B T_g, \quad (5)$$

where $P$ denotes the gas pressure, $k_B$ is the Boltzmann constant, and $T_g$ denotes the gas temperature. Additionally, the number densities of other species are calculated using Eq. (1).

The plasma chemical reactions involve both electron and ion kinetics. Electron kinetics include elastic collision, ionization, attachment, excitation, and dissociation. Their reaction rates are calculated through electron collision cross sections [49] by solving the zero-dimensional electron Boltzmann equation with the solver BOLSIG+ [50] and taken from Ref. [51]. Ion kinetics include recombination, with reaction rates taken from Ref. [52]. The source terms in Eq. (1) are determined by the net production rate, calculated based on all these reaction rates. For transport coefficients in Eq. (2), electron mobility and diffusion coefficients are calculated using BOLSIG+ [50]. Ion mobilities are taken from Ref. [49], while ion diffusion coefficients are estimated using the Einstein relation, $D_{ion} = \mu_{ion} k_B T_{ion}/e$, where $T_{ion}$ denotes the ion temperature, assumed to be equal to the gas temperature, $T_g$.

The electron energy conservation equation [Eq. (6)] is included in the LMEA framework, serving as a foundation for precise electron transport solutions, as described by Levko and Raja [53,54]:

$$\frac{\partial n_\varepsilon}{\partial t} + \nabla \cdot \mathbf{\Gamma}_\varepsilon = S_\varepsilon. \quad (6)$$

Here $n_\varepsilon$ denotes the electron energy density, defined as

$$n_\varepsilon = \frac{3}{2} n_e k_B T_e, \quad (7)$$

with $n_e$ being the electron density and $T_e$ the electron temperature; and $\mathbf{\Gamma}_\varepsilon$ denotes the electron energy flux, defined as

$$\mathbf{\Gamma}_\varepsilon = (n_\varepsilon + p_e) \mathbf{u}_e - \kappa_e \nabla T_e, \quad (8)$$

with $\mathbf{u}_e$ denoting the electron flux velocity, $\kappa_e$ the electron thermal conductivity, defined as

$$\kappa_e = \frac{5}{2} n_e k_B D_e, \quad (9)$$

and $p_e$ denoting the electron pressure, defined as

$$p_e = n_e k_B T_e. \tag{10}$$

In nonequilibrium plasmas, where $T_e \gg T_{\text{ion}}$, electrons behave as a rapidly responding fluid. In this state, the electron pressure changes slowly enough to be approximated as incompressible, resulting in $\nabla p_e = 0$. By integrating Eqs. (2) and (8)–(10), one can obtain

$$\mathbf{\Gamma}_\varepsilon = -\frac{5}{3}\mu_e n_\varepsilon \mathbf{E} - \frac{5}{3}D_e \nabla n_\varepsilon. \tag{11}$$

Thus, $\mathbf{\Gamma}_\varepsilon$ can be determined using the previously defined $\mu_e$ and $D_e$.

Finally, $S_\varepsilon$ in Eq. (6) is the source term for electron energy, defined as

$$S_\varepsilon = -e_0 \mathbf{\Gamma}_e \mathbf{E} - e_0 \sum_j \Delta\varepsilon_{\text{inel},j} R_j - \frac{3}{2}k_B n_e \frac{2m_e}{m_g}(T_e - T_g)\nu_{\text{el}}, \tag{12}$$

where $m_e$ denotes the mass of the electron, $m_g$ denotes the mass of the SF$_6$ molecule, $\nu_{\text{el}}$ denotes the frequency of elastic collisions, $\Delta\varepsilon_{\text{inel},j}$ denotes the energy change of the $j$th electron-neutral inelastic collision reaction, and $R_j$ denotes the rate of the $j$th reaction. The first term in Eq. (12) denotes the electron Joule heating, while the second and third terms denote the contributions of inelastic and elastic collisions, respectively.

The critical reduced electric field $(E/N)_{\text{cr}}$ for effective ionization calculated by BOLSIG+ [50] is approximately 360 Td, which is in good agreement with the measured benchmark reported by Christophorou and Olthoff [49]. In subsequent analyses utilizing LMEA, $(E/N)_{\text{cr}}$ is taken to be 360 Td.

#### 2. Local field approximation

In the context of the LFA, the electron energy relaxation process is neglected, reducing the difficulty of spatial numerical convergence. The LFA framework considers three species, electrons, positive ions, and negative ions, with their number densities calculated using Eq. (1). The mobilities and diffusion coefficients in Eq. (2), as well as the source terms in Eq. (1), are expressed as functions of the reduced electric field $E/N$, consistent with those in Ref. [55]. The photoionization rate $S_{\text{ph}}$ is calculated using Zhelezniak's model [56] and the Helmholtz equations for air [57,58] as an approximate alternative approach. The proportionality factor within the photo production rate is modified to 0.1.

Notably, in the context of the LFA, the critical reduced electric field $(E/N)_{\text{cr}}$ for effective ionization is 338 Td, based on ionization and attachment coefficients. It exhibits a minor deviation from the theoretical benchmark of 360 Td for pure SF$_6$, primarily due to the fitting process used for these coefficients, as performed by Morrow [55]. The fitting was based on data from multiple groups [59–66], which inevitably introduced minor deviations. Nonetheless, the fitted value remains consistent with experimental trends reported by Christophorou and Olthoff [49]. Consequently, this deviation does not fundamentally affect the underlying discharge mechanisms, and in subsequent analyses utilizing LFA, $(E/N)_{\text{cr}}$ is taken to be 338 Td.

### C. Boundary conditions

For the boundary conditions of the floating metal particle at boundary 1, the current continuity equation is used to represent the effect of plasma on particle charge,

$$\frac{\partial \sigma_s}{\partial t} = \mathbf{n} \cdot \mathbf{J}_i + \mathbf{n} \cdot \mathbf{J}_e, \tag{13}$$

where $\sigma_s$ denotes the surface-charge density and $\mathbf{n} \cdot \mathbf{J}_i$ and $\mathbf{n} \cdot \mathbf{J}_e$ denote the normal components of the total ion current density and the total electron current density on the particle surface, respectively. The equipotential condition is set for the metal particle surface, but time-dependent:

$$U_F \equiv \text{constant}, \tag{14}$$

where $U_F$ denotes the floating potential of the metal particle. Then, an integral boundary condition is set to control the overall charge $Q$ of the metal particle:

$$\int_S \mathbf{n} \cdot \mathbf{D} \, dS = Q, \tag{15}$$

where $\mathbf{n} \cdot \mathbf{D}$ denotes the normal component of the electric displacement on the particle surface.

The above settings ensure that the electric field on the metal particle surface is normal to the surface and that the entire charge on the metal particle is distributed on the surface. Finally, Poisson's equation [Eq. (3)], combined with the floating boundary conditions (14) and (15), ensure a self-consistent redistribution of the surface electric field and the surface-charge density on the floating boundary. This redistribution reflects the electrostatic induction process of the floating metal and ensures that the metal particle is in a state of electrostatic equilibrium during every computational time step.

In addition, the kinetic Maxwellian flux condition combined with a drift effect and the secondary electron emission flux, where the secondary electron emission coefficient $\gamma = 0.01$, is employed to define the boundary flux of electrons at boundary 1, boundary 3, and boundary 4. The symmetric boundary condition is employed at the axis of symmetry. The homogeneous Neumann boundary condition is employed at boundary 2. The applied voltage $U_0$ is employed at boundary 3. Finally, $U \equiv 0$ is employed at boundary 4.

### D. Initial conditions

The combined gap created by a floating particle is characterized by the presence of two distinct discharge gaps, fundamentally different from a single-gap system. This structure raises two key questions regarding the initial seed "$e$-SF$_5^+$" (pre-ionization) for the combined gap:

(1) Should the initial seed "$e$-SF$_5^+$" be introduced in both gaps simultaneously or only in one?
(2) What should be the number density of the initial seed?

The following discussions address these two questions in detail.

First, the pre-ionization is set by Gaussian distributions near both the top and bottom tips of the metal particle simultaneously:

$$n_{e,(\text{SF}_5^+)}(r,z) = n_{0,\max} \left[ \exp\left(-\frac{r^2}{2s_0^2} - \frac{(z-z_1)^2}{2s_0^2}\right) + \exp\left(-\frac{r^2}{2s_0^2} - \frac{(z-z_2)^2}{2s_0^2}\right) \right], \quad (16)$$

where $n_{0,\max}$ denotes the peak density. The parameter $s_0$ is set to 0.1 mm, and the Gaussian distribution centers, located at coordinates $(0,z_1)$ and $(0,z_2)$, are positioned 0.15 mm from the top and bottom tips of the particle, respectively. This dual-tip distribution is justified by the fact that, before the metal particle moves to the position necessary for combined gap breakdown, the field distortion at the two tips of the linear particle is strong enough to induce pre-ionization, as reported in Refs. [67–71].

Second, to accelerate the initiation and propagation of the streamer, $n_{0,\max}$ is set to $10^{13}$ m$^{-3}$. This value is artificially elevated from the typical pre-ionization level of $10^8$ m$^{-3}$, which is observed in the simulation case where $U_0 = -7$ kV, a voltage at which the gap breakdown threshold is not satisfied. A comparison of the initial stage of streamer discharge at these two pre-ionization levels is shown in Fig. 2.

At both pre-ionization levels ($n_{0,\max}$), the initial stage of discharge exhibits a similar pattern: an electron avalanche phase [Figs. 2(a) and 2(c)] followed by the inception of the streamer [Figs. 2(b) and 2(d)]. For $n_{0,\max} = 10^8$ m$^{-3}$, the electron density $n_e$ reaches $10^{18}$ m$^{-3}$ at 0.12 ns and $10^{21}$ m$^{-3}$ at 0.3 ns. At the artificially elevated level of $n_{0,\max} = 10^{13}$ m$^{-3}$, $n_e$ reaches $10^{18}$ m$^{-3}$ at 0.04 ns and $10^{21}$ m$^{-3}$ at 0.15 ns. Consequently, one can conclude that the overall characteristics of the initial stage of discharge remain consistent, and the higher pre-ionization level further significantly accelerates the development of the discharge compared to the lower level. In addition, electron detachment is also considered a kinetic mechanism to provide initial seed electrons in SF$_6$ discharges for certain scenarios [72–74]; detailed comparisons remain for further investigation in the future.

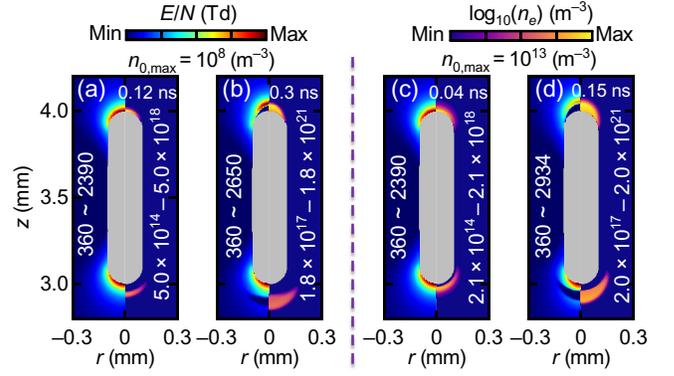

FIG. 2. Evolution of the reduced electric field $E/N$ and logarithmic electron density $\log_{10}(n_e)$ under different conditions: (a),(b) $n_{0,\max} = 10^8$ m$^{-3}$; and (c),(d) $n_{0,\max} = 10^{13}$ m$^{-3}$. For all configurations, $U_0 = -40$ kV, $Q_0 = 0$, $P = 1$ atm, and the LMEA are utilized with $(E/N)_{\text{cr}} = 360$ Td. Labels for $(E/N)_{\min}$, $(E/N)_{\max}$, $n_{e,\min}$, and $n_{e,\max}$ are shown in each panel.

The gas temperature $T_g$ is set to 300 K, and the gas pressure $P$ varies according to the specific studies discussed below. These two parameters are assumed to remain constant throughout the entire streamer process.

## III. INCEPTION OF DOUBLE-END STREAMER

Under the condition $Q_0 = 0$, streamer inception induced by a floating linear metal particle occurs simultaneously at both ends, as shown in Figs. 2(b) and 2(d), differing notably from the behavior observed in floating dielectrics and microdischarge theory. Mirpour and Nijdam [75] reported that, for floating dielectric (TiO$_2$) particles under negative voltage, the positive streamer initiated first due to its lower threshold, while the negative streamer initiated only after the positive streamer had broken down. Microdischarge theory [12–16] assumed that short-gap breakdown occurred first to establish electrical connection, allowing the floating metal particle to adopt the electrode potential, which then triggered subsequent long-gap discharge. In contrast, this paper finds that streamer inception in the long gap does not rely on short-gap breakdown, namely, a double-end streamer consistently occurs in the context of floating-metal-particle-induced discharge. The detailed mechanisms are discussed below.

First, as shown in Fig. 3, under the initial condition $Q_0 = 0$, the electric fields at both tips of the linear metal particle are approximately equal, regardless of the applied voltage. This ensures that both tips reach the effective ionization threshold $(E/N)_{\text{cr}} = 360$ Td simultaneously, enabling ionization at both tips, as shown in Figs. 2(a) and 2(c). However, ionization alone does not necessarily initiate a streamer. The electrostatic induction of the metal particle plays a critical role in the occurrence of the double-end streamer.

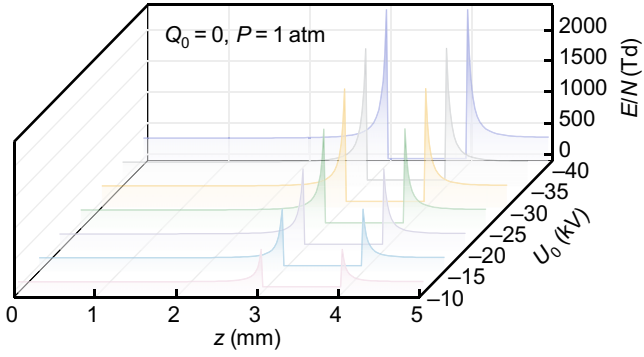

FIG. 3. Distribution of the reduced electric field $E/N$ along the axis of symmetry at the initial moment ($t = 0$) under different $U_0$.

Further analysis of the simulation case with $U_0 = -40$ kV, $Q_0 = 0$, and $P = 1$ atm illustrates this process, as shown in Fig. 4. During the early stage of ionization ($\sim 0.1$ ns), the electron conduction current at the top tip significantly exceeds the positive-ion conduction current at the bottom tip due to the higher mobility of electrons [see Fig. 4(a)]. As a result, at 0.1 ns, the total charge $Q$ on the metal particle reaches $-15$ pC. To maintain electrostatic equilibrium, this additional negative charge is redistributed, causing the surface charge at the bottom tip to increase from $-1035$ to $-1127$ pC mm$^{-2}$ [see Fig. 4(b)] in such a short period. According to Gauss's law, the surface-charge density of the metal particle satisfies $\sigma_s = \varepsilon E_s$. As a result, during the initial stage of ionization ($\sim 0.1$ ns), the electric field at the bottom tip, already exceeding $(E/N)_{\text{cr}}$, is further increased by $\sim 200$ Td. This enhancement promotes the inception of the double-end streamer, as shown in Fig. 2.

## IV. ROLE OF METAL PARTICLES: THE FOLLOWING-STREAMER MECHANISM

The breakdown mechanism in the combined gas gap induced by floating metal particles primarily highlights how these particles facilitate long-gap ($g_l$) breakdown. This section investigates the physical processes through which floating metal particles contribute to long-gap breakdown. As shown in Fig. 5(i), the primary streamer in the long-gap breakdown is followed by two independent streamers. The space-charge field of following streamers exceeds 20% of the background electric field; thus they are considered to be newly generated streamers. From a phenomenological perspective, floating metal particles play a pivotal role, which we refer to in this paper as the "following-streamer" mechanism. This pattern closely resembles the multiple-streamer events commonly observed under pulsed voltage conditions [76–79].

Notably, it is essential to distinguish the subsequent streamers discussed here from those arising solely from the accumulation of negative ions in SF$_6$ gas, as described in Ref. [18]. The primary difference lies in the characteristics of the electric field within the streamer channel. For the secondary streamer described in Ref. [18], the channel field is quickly shielded below the critical value $(E/N)_{\text{cr}}$. In contrast, for the FS mechanism discussed in this paper, the channel field remains above $(E/N)_{\text{cr}}$ for a period after the streamer is initiated, as shown in Figs. 5(b) and 5(i).

From a physical perspective, this difference can be attributed to the negative surface charge at the bottom tip of the floating metal particle, which primarily drives the multiple streamers in the FS mechanism, as shown in Fig. 5(g). Because the metal particle remains in electrostatic equilibrium, Gauss's law ensures that the condition $\sigma_s = \varepsilon E_s$ is consistently satisfied on its surface. This equilibrium causes the surface electric field at the particle's bottom tip to intensify, similar to the effect of the rising edge of a pulsed voltage. As a result, the linear floating metal particle induces subsequent streamers referred to as FS. The detailed physical mechanisms underlying how the metal particle induces FS are further discussed in Sec. IV A.

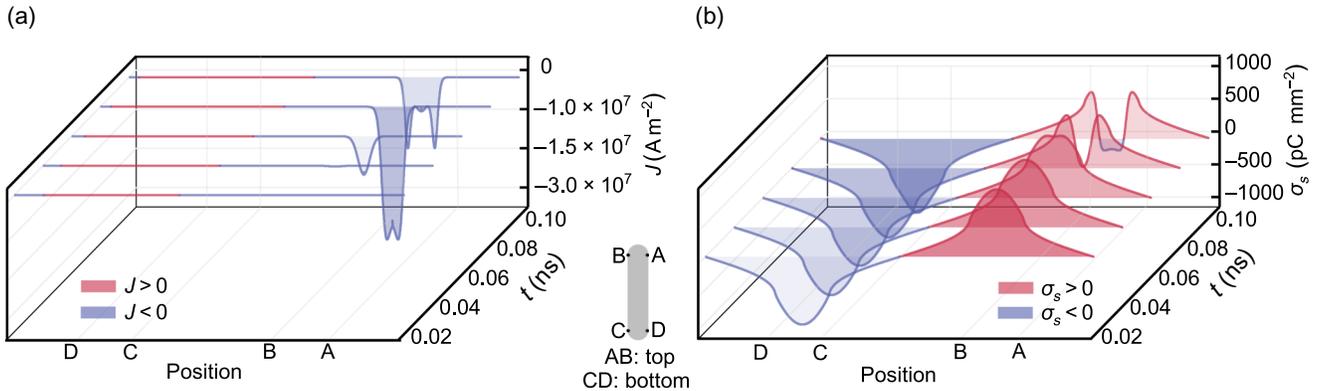

FIG. 4. Evolution of (a) current density $J$ and (b) surface-charge density $\sigma_s$ along the surface of the linear metal particle, illustrating the inception of the double-end streamer.

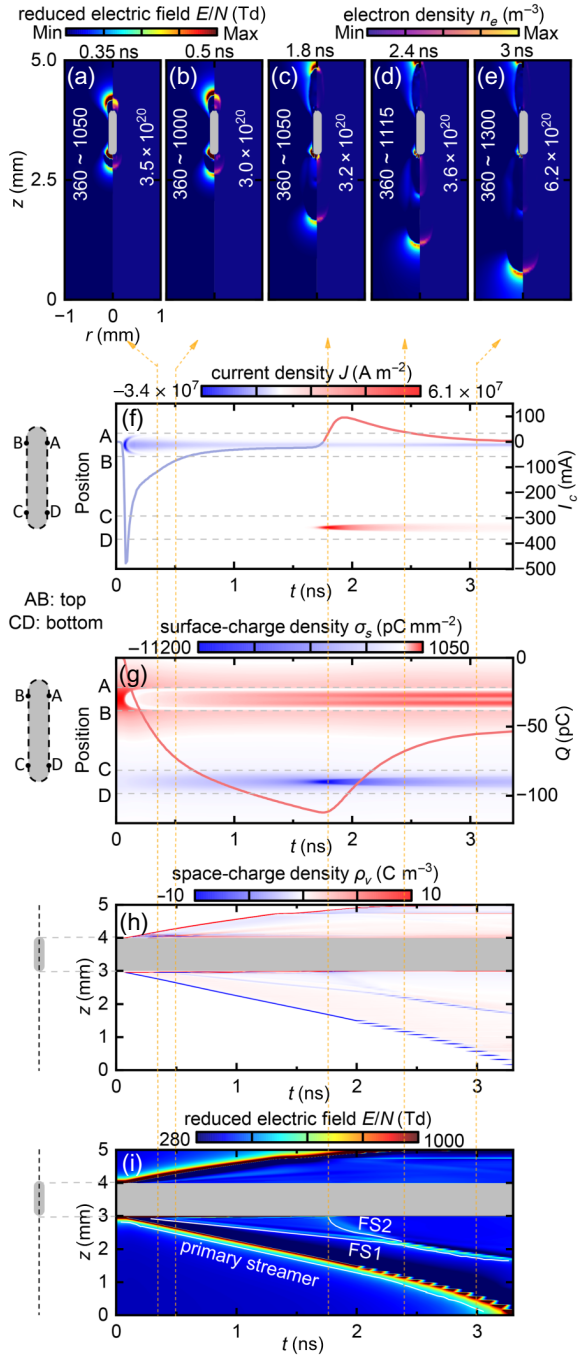

FIG. 5. (a)–(e) Evolution of the reduced electric field $E/N$ and electron density $n_e$. Labels for $(E/N)_{min}$, $(E/N)_{max}$, and $n_{e,max}$ are shown in each panel, where $n_{e,min}$ is fixed at 0. (f) Spatiotemporal evolution of current density $J$ along the surface of a metal particle and the temporal evolution of the total conduction current $I_c$. (g) Spatiotemporal evolution of surface-charge density $\sigma_s$ along the surface of a metal particle and the temporal evolution of the total charge $Q$. (h) Spatiotemporal evolution of space-charge density $\rho_v$ along the axis of symmetry. (i) Spatiotemporal evolution of the reduced electric field $E/N$ along the axis of symmetry. The white lines show the position of the streamer head. In all panels, $U_0 = -40$ kV, $Q_0 = 0$, $P = 1$ atm, and the LMEA is utilized with $(E/N)_{cr} = 360$ Td.

### A. Dynamics and mechanism of following streamers

At 0.5 ns, the first following streamer (FS1) initiates, as shown in Figs. 5(a) and 5(b). This process is driven by two critical aspects: the electron charging effect at the particle's top tip and the electrostatic induction of the entire metal particle, both of which are indispensable. During the 0–0.5 ns interval, under the influence of the negative applied voltage, electrons migrate rapidly toward the particle's top tip, generating a negative conduction current at this location. In contrast, ions migrate more slowly, and no significant conduction current is observed at the particle's bottom tip. As a result, the total conduction current $I_c$ acting on the metal particle peaks at $-500$ mA [see Fig. 5(f)], highlighting the dominant role of electron charging during this phase. This charging process causes a steady accumulation of negative charge $Q$ on the metal particle, exceeding $-73$ pC, as shown in Fig. 5(g).

Notably, although the negative charge primarily accumulates at the particle's top tip due to electron conduction, the negative surface charge at the bottom tip also increases significantly, rising from $-1034$ pC mm$^{-2}$ at 0 ns to $-2076$ pC mm$^{-2}$ at 0.5 ns [see Fig. 5(g)]. This behavior is fundamentally governed by the electrostatic induction of the floating metal particle; namely, to maintain electrostatic equilibrium, the surface charge on the particle undergoes a self-consistent redistribution. As a result, the negative surface charge gradually increases at the particle's bottom tip. The Coulomb electric field induced by this negative surface charge, aligned with the applied electric field, enhances the local electric field near the particle's bottom tip, thereby enhancing ionization in the sheath in this region, as shown in Fig. 5(a).

The positive feedback mechanism described in Ref. [18] further promotes the attachment of sheath electrons near the bottom tip, resulting in the formation of a negative-ion region and an increase in negative space charge within this region [see Fig. 5(h)]. The combined effect of the negative space charge in the negative-ion region and the negative surface charge at the particle's bottom tip intensifies the local electric field. After 0.35 ns, the reduced field outside the sheath exceeds the critical value $(E/N)_{cr}$. By 0.5 ns, FS1 initiates from this position and propagates forward, maintaining the same polarity as the primary streamer, with both being negative streamers [see Fig. 5(h)].

At 1.8 ns, the second following streamer (FS2) initiates, as shown in Figs. 5(c) and 5(d). This phenomenon is driven by two aspects: the positive-ion charging effect at the particle's bottom tip and the electrostatic induction of the entire metal particle. Starting at 1.72 ns, ion migration becomes significant. The particle's bottom tip sustains a positive conduction current (mainly from positive ions), while the top tip sustains a negative conduction current (dominated by electrons, which contribute six times more than negative ions). After 1.8 ns, the positive conduction current at the bottom tip surpasses the negative conduction current at the top tip, causing the total conduction current $I_c$ to exceed 65 mA [see Fig. 5(f)], and leading to a decrease in the net negative charge $Q$ on the particle [see Fig. 5(g)].

Notably, although the positive charge primarily accumulates at the bottom tip due to positive-ion conduction, electrostatic induction redistributes some of this positive charge to the top tip, increasing the positive surface charge there. Meanwhile, the bottom tip induces more negative surface charge, reaching a maximum density of $-11\,200$ pC mm$^{-2}$. The negative surface charge at the bottom tip, together with the negative space charge near the particle's bottom tip and the positive surface charge at the top tip, establish a self-consistent electrostatic equilibrium. At this stage, the induced negative surface charge at the bottom tip significantly enhances the local electric field near the bottom tip. The combined effects of the negative space charge in the negative-ion region and the negative surface charge ultimately induce the second following streamer (FS2), which remains the same negative polarity as the previous streamers, as shown in Fig. 5(h).

### B. Following-streamer facilitation on breakdown

The propagation speed of the streamer is qualitatively estimated from the slope of the streamer head trajectory line in Fig. 5(i), namely, a steeper slope (greater displacement in the $z$ direction per unit time) indicates faster propagation.

The facilitating effect of FS1 on long-gap breakdown is primarily driven by the space-charge field generated by the negative space charge at its head. As shown in Fig. 5(h), FS1, functioning as a negative streamer, generates the space-charge field that is aligned with the applied electric field. This alignment enhances the local field at the primary streamer head, maintaining the reduced electric field at $\sim$1000 Td, as shown in Fig. 5(i). In comparison to the single-negative-streamer scenario reported in Ref. [80], where the electric field at the streamer head typically decreases by $\sim$30 Td and the propagation speed drops by 43%, the presence of FS1 in this paper mitigates the decline of the electric field at the primary streamer head. As a result, under the influence of FS1, the propagation speed of the primary streamer remains nearly constant, as shown in Fig. 5(i), instead of slowing down. Thus, one can conclude that FS1 significantly enhances the propagation of the primary streamer, thereby facilitating the breakdown of the long gap.

The facilitating effect of FS2 on long-gap breakdown arises from its merging with FS1. As shown in Fig. 5(h), at 2.4 ns, FS2 merges with FS1, causing an increase in the space-charge density $\rho_v$ at the head of FS1 from $-1.5$ C m$^{-3}$ at 1.8 ns to $-5.3$ C m$^{-3}$ at 2.4 ns. This increase in space charge strengthens the electric field at the head of FS1, raising it from 480 Td at 1.8 ns to 708 Td at 2.4 ns, as shown in Fig. 5(i). After the merging, the propagation speed of FS1 is significantly accelerated. The rise in negative space charge at the head of FS1 also enhances the reduced field at the primary streamer head, increasing it from 1050 Td at 1.8 ns to 1115 Td at 2.4 ns, as shown in Figs. 5(c) and 5(d). Consequently, after 2.4 ns, the propagation speed of the primary streamer noticeably accelerates, as shown in Fig. 5(i). Thus, one can conclude that FS2 significantly enhances the propagation of the primary streamer, ultimately facilitating the breakdown of the long gap.

### C. Preliminary study of the initial charge $Q_0$

To investigate the effect of initial particle charge on discharge behavior, $Q_0$ is set to $+50$, $+100$, and $-100$ pC in the initial conditions, while all the other parameters remain unchanged. Preliminary results indicate that double-end streamer inception and the formation of following streamers still occur under these $Q_0$ conditions. However, the different $Q_0$ values significantly influence the propagation speed of streamers at both ends, as well as the number of following streamers. A detailed discussion of these cases and suggestions for future work are presented in Appendix B.

## V. NONLINEAR BEHAVIOR OF HIGH-PRESSURE SF$_6$: SIDE-STREAMER MECHANISM

In practical applications, gas-insulated electrical equipment typically operates at pressures above atmospheric levels. Consequently, particle-induced SF$_6$ discharge in real scenarios inevitably occurs under high-pressure conditions. Therefore, we investigate the effect of gas pressure on SF$_6$ streamer propagation in the context of particle-structure electrodes. The use of particle-structure electrodes is justified by two main considerations. First, it aligns with the subject of this paper, ensuring consistency in the investigative framework. Second, analyzing the effects of gas pressure within the particle-structure framework is physically valid. Specifically, gas pressure does not fundamentally alter the interaction between the metal particle and the space charge, as this interaction is governed by the intrinsic electrostatic induction properties of the metal. This decoupling of gas pressure effects from the behavior of the metal particle confirms that the particle-structure electrode is adequate for addressing the research objectives.

As outlined in Sec. I, the nonlinear breakdown behavior of high-pressure SF$_6$ is closely linked to the distinctive streamer morphologies under high-pressure conditions. In this section, we investigate the dynamics of high-pressure

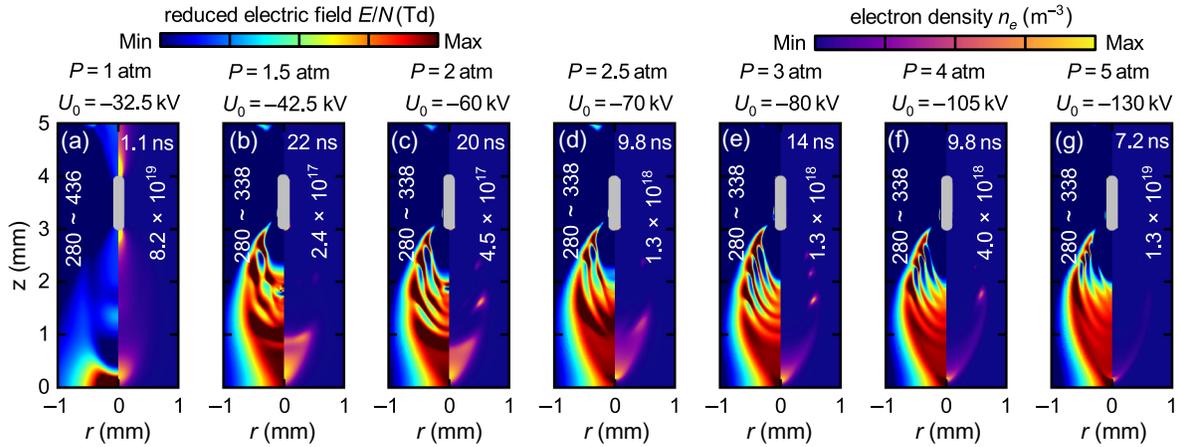

FIG. 6. Evolution of the reduced electric field $E/N$ and electron density $n_e$ under different conditions: (a) $P = 1$ atm, $U_0 = -32.5$ kV; (b) $P = 1.5$ atm, $U_0 = -42.5$ kV; (c) $P = 2$ atm, $U_0 = -60$ kV; (d) $P = 2.5$ atm, $U_0 = -70$ kV; (e) $P = 3$ atm, $U_0 = -80$ kV; (f) $P = 4$ atm, $U_0 = -105$ kV; and (g) $P = 5$ atm, $U_0 = -130$ kV. Herein, $Q_0 = 0$ and the LFA is utilized with $(E/N)_{cr} = 338$ Td for all configurations. Labels for $(E/N)_{min}$, $(E/N)_{max}$, and $n_{e,max}$ are shown in each panel, where $n_{e,min}$ is fixed at 0.

SF$_6$ streamer morphologies in the context of the LFA framework. It should be acknowledged that, although the LFA framework cannot fully capture the interaction between the space charge and the metal particle, it is effective for modeling solely the dynamics of streamer propagation in the combined gap, as reported in Refs. [81–83]. This makes the LFA an appropriate framework for the objectives of this section, and it also offers the advantage of significantly reducing computational costs. For simplicity, in the following discussions, the term "streamer" will exclusively refer to those occurring in the long gap, excluding those in the short gap.

### A. Dynamics of side streamers

#### 1. Side-streamer phenomenon

The simulation cases are conducted at pressure levels of 1, 1.5, 2, 2.5, 3, 4, and 5 atm. In all the cases below, the initial charge $Q_0 = 0$. The applied voltage $U_0$ is set at a near-critical voltage, exceeding the breakdown threshold. This voltage allows for an observation qualitatively consistent with critical behavior while also highlighting the characteristics of streamer propagation. The results of SF$_6$ streamer propagation at near-critical voltages under varying pressures are shown in Fig. 6. For clarity, the term "primary streamer" refers to the discharge at the forefront, propagating along the axis.

At 1 atm, as shown in Fig. 6(a), the streamer behavior is consistent with the findings in Sec. IV, exhibiting multiple streamer characteristics. In contrast, at pressures of 1.5 atm or higher, as shown in Figs. 6(b)–6(g), the streamer develops a distinct coherent structure, markedly different from the patterns observed at atmospheric pressure. This coherent structure forms along the side of the primary streamer channel and is characterized by a localized region of field enhancement where $E/N > (E/N)_{cr}$. Within this enhanced field region, localized areas of high electron density $n_e$, exceeding $10^{17}$ m$^{-3}$, indicate the occurrence of intense ionization. To determine whether this coherent structure constitutes the formation of a new streamer, it is necessary to determine if the field enhancement regions within it shift spatially, which could signal the propagation of an ionization wave.

For detailed analysis, the simulation case with $P = 4$ atm and $U_0 = -105$ kV is examined, as shown in Fig. 7, which illustrates the complete streamer propagation process. A coherent structure is observed forming along the side of the primary streamer and propagating forward as an ionization wave in the same direction, as shown in Figs. 7(b)–7(e). This coherent structure essentially behaves as a new streamer with the same polarity as the primary one. In this paper, we refer to this new type of streamer as a "side streamer" to emphasize its formation and propagation alongside the primary streamer. Interestingly, multiple SS can emerge during a single discharge event, as shown in Figs. 7(f) and 7(g). Each SS follows a fundamentally similar pattern, forming sequentially along the side of its predecessor. This process results in a characteristic "layer-by-layer nesting" spatial arrangement.

#### 2. Impact of side streamer on primary streamer

Another important aspect to investigate is the impact of the SS phenomenon on the primary streamer. In the presence of SS, the primary streamer exhibits two distinct propagation modes. The first mode, referred to as "primary intermittency," is shown in Figs. 7(b)–7(e). In this mode, during SS propagation, the reduced field $E/N$ at the primary streamer head gradually decreases to below $(E/N)_{cr}$, accompanied by a reduction in electron density $n_e$ by approximately one order of magnitude. This is due to the Laplace electric field being $\sim$218 Td at positions far from the tip, which is insufficient to maintain effective ionization at the streamer head. Specifically, the effective ionization rate decreases to $10^{29}$ m$^{-3}$ s$^{-1}$ at 0.8 ns and becomes negative by 1.2 ns. This temporarily halts the self-maintained propagation of the primary streamer, as shown in Figs. 7(b) and 7(c). However, when the SS reaches the primary streamer head, the head field $E/N$ recovers to above $(E/N)_{\rm cr}$, re-establishing intense ionization. This recovery increases the electron density $n_e$ at the streamer head from $2 \times 10^{18}$ to $7.8 \times 10^{18}$ m$^{-3}$, as shown in Figs. 7(d) and 7(e). These characteristics demonstrate a transition in the primary streamer from a temporary pause to resumed axial propagation, defining the primary intermittency mode.

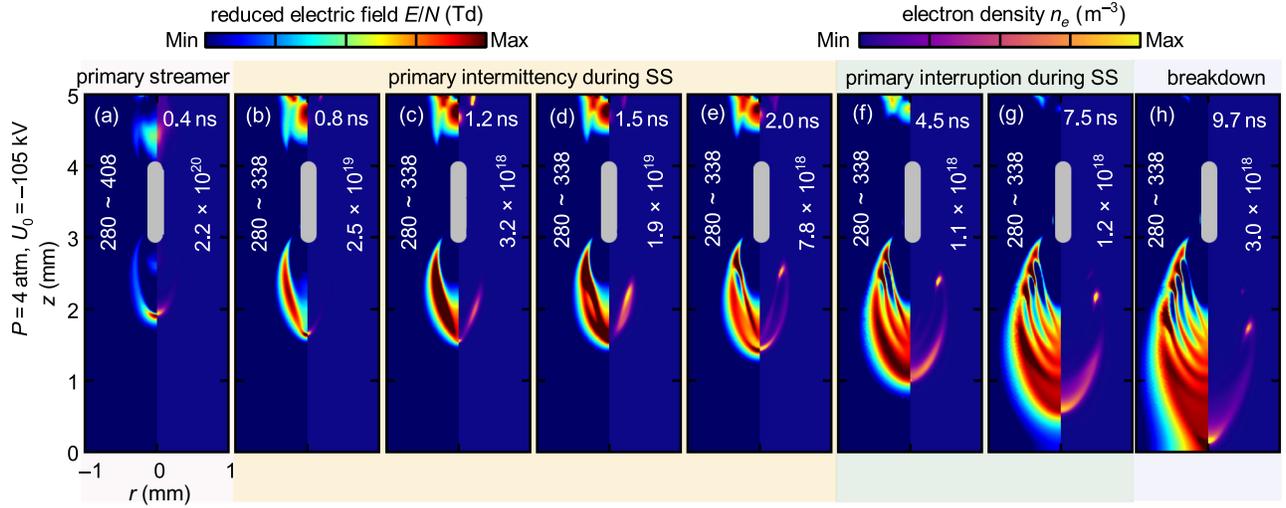

FIG. 7. Evolution of the reduced electric field $E/N$ and electron density $n_e$ under near-critical voltage $U_0 = -105$ kV and at a pressure of 4 atm. Herein, $Q_0 = 0$ and the LFA is utilized with $(E/N)_{\rm cr} = 338$ Td for all configurations. Labels for $(E/N)_{\rm min}$, $(E/N)_{\rm max}$, and $n_{e,\rm max}$ are shown in each panel, where $n_{e,\rm min}$ is fixed at 0.

After propagating a certain distance, the primary streamer transitions to another mode, referred to as "primary interruption." In this mode, even after the SS merges with the primary streamer, the streamer head does not fully recover its strong reduced field, remaining at $E/N < (E/N)_{\rm cr}$, as shown in Figs. 7(f) and 7(g). This indicates that the primary streamer remains in an interrupted state, with insufficient ionization level at its head. Surprisingly, an unusual phenomenon is observed: despite this interruption, the region ahead of the primary streamer, characterized by an enhanced but subcritical reduced field $E/N < (E/N)_{\rm cr}$, continues to propagate forward. In this forward region, the electron density remains around $n_e \approx 1 \times 10^{18}$ m$^{-3}$. Although this density is lower than that during fully active streamer propagation, it reflects the continuous generation of new electrons through ionization.

These observations indicate that, under high-pressure conditions, SF$_6$ streamer propagation is strongly influenced by the SS phenomenon. The presence of SS leads the primary streamer to exhibit distinct propagation modes. The underlying physical mechanisms driving the SS phenomenon and its influence on the primary streamer are analyzed in the following sections.

### B. Side-streamer mechanism and its facilitation on breakdown

To investigate the physical mechanisms underlying the SS phenomenon, the simulation case with $P = 4$ atm and $U_0 = -105$ kV is analyzed in detail in this section. Given the uniform behavior of each SS within the layer-by-layer nesting pattern, the first SS event is selected as a representative case. Key physical parameters, including the photoionization rate ($S_{\rm ph}$), effective ionization rate ($S_{\rm eff}$), space-charge density ($\rho_v$), and reduced electric field ($E/N$), are systematically examined. For clarity and precision, the SS phenomenon is divided into four distinct stages: pre-initiation, initiation, propagation, and merging. Each stage is individually analyzed to provide a comprehensive understanding of the underlying mechanisms.

The pre-initiation stage is analyzed at 0.75 and 0.81 ns to illustrate the preparatory processes leading to SS formation. As shown in Figs. 8(a1), 8(a2), 8(b1), and 8(b2), on the side of the primary streamer, the effective ionization rate ($S_{\rm eff} < 0$) and photoionization rate ($S_{\rm ph} > 0$) exhibit opposite behaviors due to their different physical definitions. Specifically, $S_{\rm ph}$ is positively correlated with $\alpha n_e \mu_e E$, where $\alpha$ denotes the ionization coefficient, while $S_{\rm eff} = (\alpha - \eta) n_e \mu_e E$, where $\eta$ denotes the attachment coefficient. Thus, $S_{\rm ph}$ is solely related to collision ionization, while $S_{\rm eff}$ reflects the net effect of ionization and attachment during the electron collision process. This distinction results in continuous generation of seed electrons on the side of the channel via photoionization, even as they are rapidly attached, leading to the accumulation of negative ions there. As a result, the negative space charge on the side increases, as shown in Figs. 8(a3) and 8(b3). The space-charge field generated aligns with the applied negative-polarity electric field, enhancing the local field on the side from 334 to 338 Td [$(E/N)_{cr}$], as shown in Figs. 8(a4) and 8(b4).

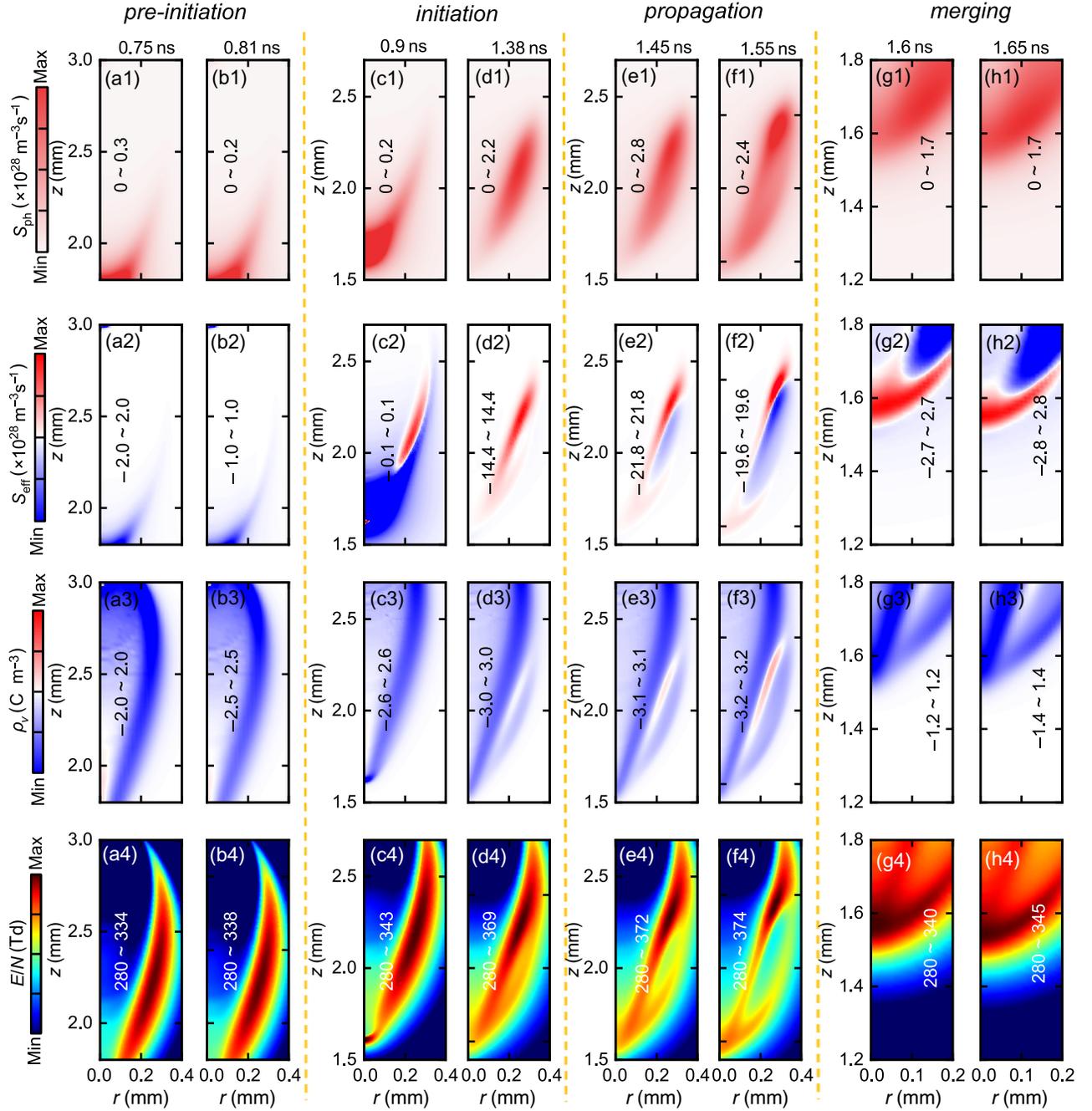

FIG. 8. Evolution of the (a1)–(h1) photoionization rate $S_{ph}$, (a2)–(h2) effective ionization rate $S_{eff}$, (a3)–(h3) space-charge density $\rho_v$, and (a4)–(h4) reduced electric field $E/N$, illustrating the SS mechanism under near-critical voltage $U_0 = -105$ kV and at a pressure of 4 atm. Herein, $Q_0 = 0$ and the LFA is utilized with $(E/N)_{cr} = 338$ Td for all configurations.

The initiation stage is analyzed at 0.9 and 1.38 ns to illustrate the onset of SS formation. At 0.9 ns, the continued accumulation of negative ions from the pre-initiation stage causes the local field on the side to exceed $(E/N)_{cr}$, as shown in Fig. 8(c4). This marks the beginning of electron generation through the electron collision

ZIHAO FENG *et al.*process, although it remains weaker than photoionization ($S_{\text{eff}} < S_{\text{ph}}$), as shown in Figs. 8(c1) and 8(c2). At 1.38 ns, a spatial shift in the space-charge distribution signals the emergence of a new ionization wave (SS) with the same negative polarity as the primary streamer, as shown in Fig. 8(d3). During this time, $S_{\text{eff}} \approx 6.5 S_{\text{ph}}$, as shown in Figs. 8(d1) and 8(d2), indicating a significant intensification of the electron collision process. This intensification allows the electron collision process to surpass photoionization as the dominant electron generation mechanism, thereby facilitating the formation of the SS.

The propagation stage is analyzed at 1.45 and 1.55 ns to illustrate the behavior of SS as an ionization wave. During this stage, the effective ionization rate ($S_{\text{eff}}$) significantly exceeds the photoionization rate ($S_{\text{ph}}$), with $S_{\text{eff}} \approx 8 S_{\text{ph}}$. This indicates that the electron collision process dominates electron generation. As the SS propagates forward, the negative space charge at its head generates a shielding effect that weakens the electric field in the region behind it, reducing the reduced field below $(E/N)_{\text{cr}}$, as shown in Figs. 8(e4) and 8(f4). In this low-field region, electron attachment surpasses electron generation, resulting in $S_{\text{eff}} < 0$. This process leads to the formation of a well-defined streamer channel of SS.

The merging stage is analyzed at 1.6 and 1.65 ns to illustrate the process of the SS merging with the primary streamer and its facilitation on breakdown. As the SS propagates forward, it gradually moves toward the axis where the primary streamer is located. This movement is driven by the negative space charge generated by the primary streamer, as shown in Figs. 8(g3) and 8(h3). The negative space charge enhances the local field near the axis, as shown in Figs. 8(g4) and 8(h4), which in turn intensifies ionization near the axis, as shown in Figs. 8(g2) and 8(h2). As a result, the SS develops consistently toward the axis until it merges with the primary streamer. Following the merging, the SS disappears, and the negative space charge on the axis increases from $-1.2$ to $-1.4$ C m$^{-3}$, as shown in Figs. 8(g3) and 8(h3). This increase in space charge restores the local electric field at the primary streamer head to 345 Td, exceeding the critical value $(E/N)_{\text{cr}}$. As a result, the primary streamer resumes its propagation along the axis, forming the primary intermittency mode as described in Sec. V A 2.

After the merging stage of the SS, the primary streamer can exhibit an unusual phenomenon: the primary interruption mode, as shown in Figs. 7(f) and 7(g) and described in Sec. V A 2. The primary physical factor driving its propagation is photoionization, with the detailed mechanism illustrated in Fig. 9. The SS merging restores the primary streamer head to a near-critical field strength, i.e., near but below the critical value $(E/N)_{\text{cr}}$ [see Fig. 9(a4)]. Thus, photoionization remains consistently active [see Figs. 9(a1)–9(d1)] and the effective ionization rate $S_{\text{eff}}$ remains negative [see Figs. 9(a2)–9(d2)]. As a result, the seed electrons at the front generated by photoionization rapidly attach to form negative ions at the front, gradually expanding the negative space-charge region, as shown in Figs. 9(c3)–9(d3). This negative space charge enhances the local field at the front, enabling the near-critical enhanced field region to expand forward spatially, even though the field remains below $(E/N)_{\text{cr}}$, as shown in Figs. 9(c4)–9(d4).

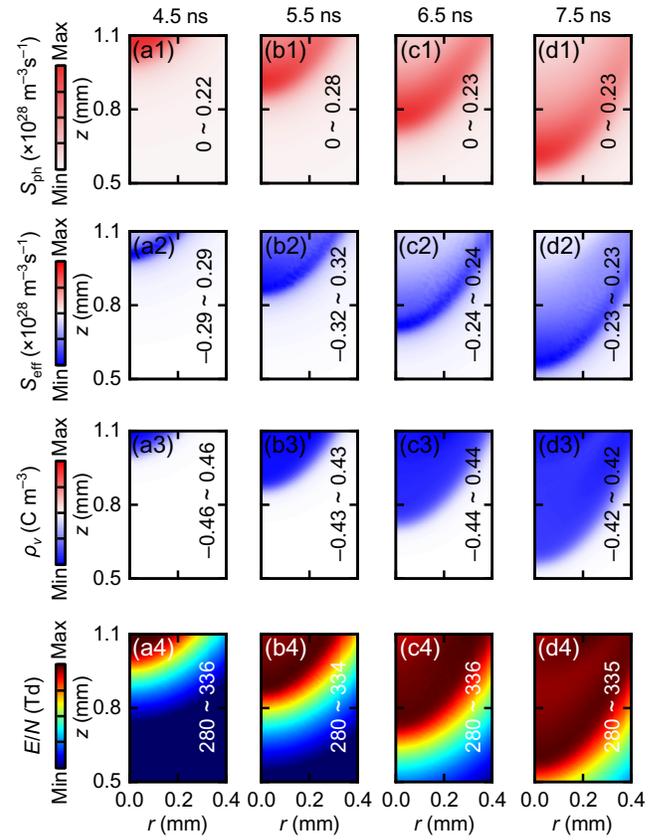

FIG. 9. Evolution of the (a1)–(d1) photoionization rate $S_{\text{ph}}$, (a2)–(d2) effective ionization rate $S_{\text{eff}}$, (a3)–(d3) space-charge density $\rho_v$, and (a4)–(d4) reduced electric field $E/N$, illustrating the primary interruption mode under near-critical voltage $U_0 = -105$ kV and at a pressure of 4 atm. Herein, $Q_0 = 0$ and the LFA is utilized with $(E/N)_{\text{cr}} = 338$ Td for all configurations.

In summary, the primary interruption mode is based on the SS merging, which restores the primary streamer head to a near-critical field strength, maintaining a relatively high level of photoionization. On this basis, the collaboration of photoionization and attachment drives the forward expansion of the negative space-charge region at the streamer head, leading to its propagation. As it expands further and approaches the vicinity of the grounded electrode, the field strength at the head is enhanced once again, exceeding $(E/N)_{\text{cr}}$, as shown in Fig. 7(h). This enables the primary streamer to continue propagating, ultimately resulting in breakdown. The charge accumulated on the metal particle at the end of this side streamer process is $-26.3$ pC.

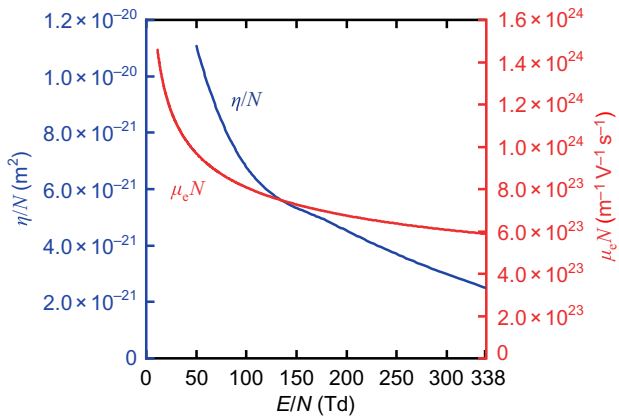

FIG. 10. Variations of $\eta/N$ and $\mu_e N$ as functions of $E/N$.

### C. Pressure dependence of side streamers: Qualitative analysis

The reduced electric field $E/N$ at the SS initiation position remains lower than the critical value $(E/N)_{cr}$ during the pre-initiation stage, as shown in Fig. 8. Therefore, the subsequent analysis is confined to the range $E/N < (E/N)_{cr}$. Since this section focuses on the dependence of gas pressure, the applied electric field $E_0$ is considered the same across different pressure conditions.

As discussed in Sec. V B, the primary species driving SS initiation is the negative ion, and its production rate $S_n$ is defined as $S_n = \eta \mu_e n_e E = (\eta/N)(\mu_e N) n_e E$. Based on the expressions for $\eta/N$ and $\mu_e N$ in Ref. [55], their variations as functions of $E/N$ are shown in Fig. 10. It is important to note that, during the early pre-initiation stage of SS, the initiation position retains the properties of the SF$_6$ streamer channel. According to Refs. [18], the electric field $E$ within the SF$_6$ streamer channel consistently recovers to a level roughly equivalent to the applied electric field $E_0$. Moreover, the electron density $n_e$ at the SS initiation position, approximately $n_e \approx 10^{17}$ m$^{-3}$, is supplied by photoionization and varies little under different pressure conditions. Consequently, the initial value of $S_n$ under different pressures, which determines the initiation of SS, is primarily determined by $\eta/N$ and $\mu_e N$.

From a qualitative perspective, as $P$ increases, $N$ increases; since $E = E_0$ remains constant, $E/N$ decreases. Both $\eta/N$ and $\mu_e N$ are monotonically decreasing functions of $E/N$, as shown in Fig. 10, thus $(\eta/N)(\mu_e N)$ increases. Given that $S_n \propto (\eta/N)(\mu_e N)$, it follows that $S_n$ increases. Hence, one can conclude that higher pressures facilitate negative-ion generation, making SS initiation more likely, as shown in Fig. 6.

The critical pressure for SS formation identified in this study (1.5 atm) falls within the pressure range (1.2–2 atm) previously reported as the inflection point on the SF$_6$ nonlinear breakdown voltage curve (U-P curve) in some experimental studies [30,84]. Additionally, relating experiments under high pressure have observed the development of side discharge channels, such as precursors [27–29] and side spark paths [30].

### D. Dynamics of side streamers under overvoltage

In practical gas-insulated equipment, overvoltage conditions are inevitable. Therefore, it is essential to examine whether the SS mechanism continues to influence SF$_6$ high-pressure streamer breakdown under overvoltage conditions. This section investigates whether the role of the SS mechanism in the development of a primary streamer remains consistent at overvoltage levels compared to its behavior at near-critical voltage. To address this, simulation cases are conducted at a representative gas pressure of 4 atm. The applied voltages $U_0$ are varied as $-110$, $-115$, and $-120$ kV, respectively. The corresponding results are shown in Fig. 11.

When the applied voltage is set to $U_0 = -110$ kV, corresponding to a background reduced field $(E/N)_b \approx 225$ Td, the SS phenomenon remains observable and the primary streamer propagates in the same primary intermittency mode as observed at near-critical voltage. This indicates that, at this overvoltage level, the elevated $(E/N)_b$ enables the local field at the head of the primary streamer to recover above $(E/N)_{cr}$ after merging with the SS, as shown in Figs. 11(a5)–11(a7). However, despite the elevated $(E/N)_b$ compared to the near-critical condition, the primary streamer still experiences a temporary pause before merging with the SS, as shown in Figs. 11(a2)–11(a4). This indicates that, at this overvoltage level, the elevated $(E/N)_b$ alone is insufficient to sustain continuous propagation. Instead, the additional negative space charge contributed by the merging SS remains the key factor that enables the primary streamer to resume propagation.

When the applied voltage is increased to $U_0 = -115$ kV, corresponding to a background reduced field $(E/N)_b \approx 235$ Td, the SS phenomenon remains observable; however, the primary streamer exhibits a new propagation mode, referred to as "primary continuous propagation," distinct from the primary intermittency or primary interruption modes discussed in Sec. V A 2. As shown in Figs. 11(b1)–11(b8), the local field at the primary streamer head consistently exceeds $(E/N)_{cr}$ throughout propagation, and the electron density $n_e$ at the head remains greater than $10^{19}$ m$^{-3}$. This behavior results from the elevated $(E/N)_b$, which enables a high ionization level, allowing the primary streamer to sustain independent propagation without relying on additional negative space charge from SS merging. Nevertheless, the SS phenomenon continues to play a supportive role. During merging, the head field is further enhanced, increasing from 357 to 368 Td, facilitating faster propagation.

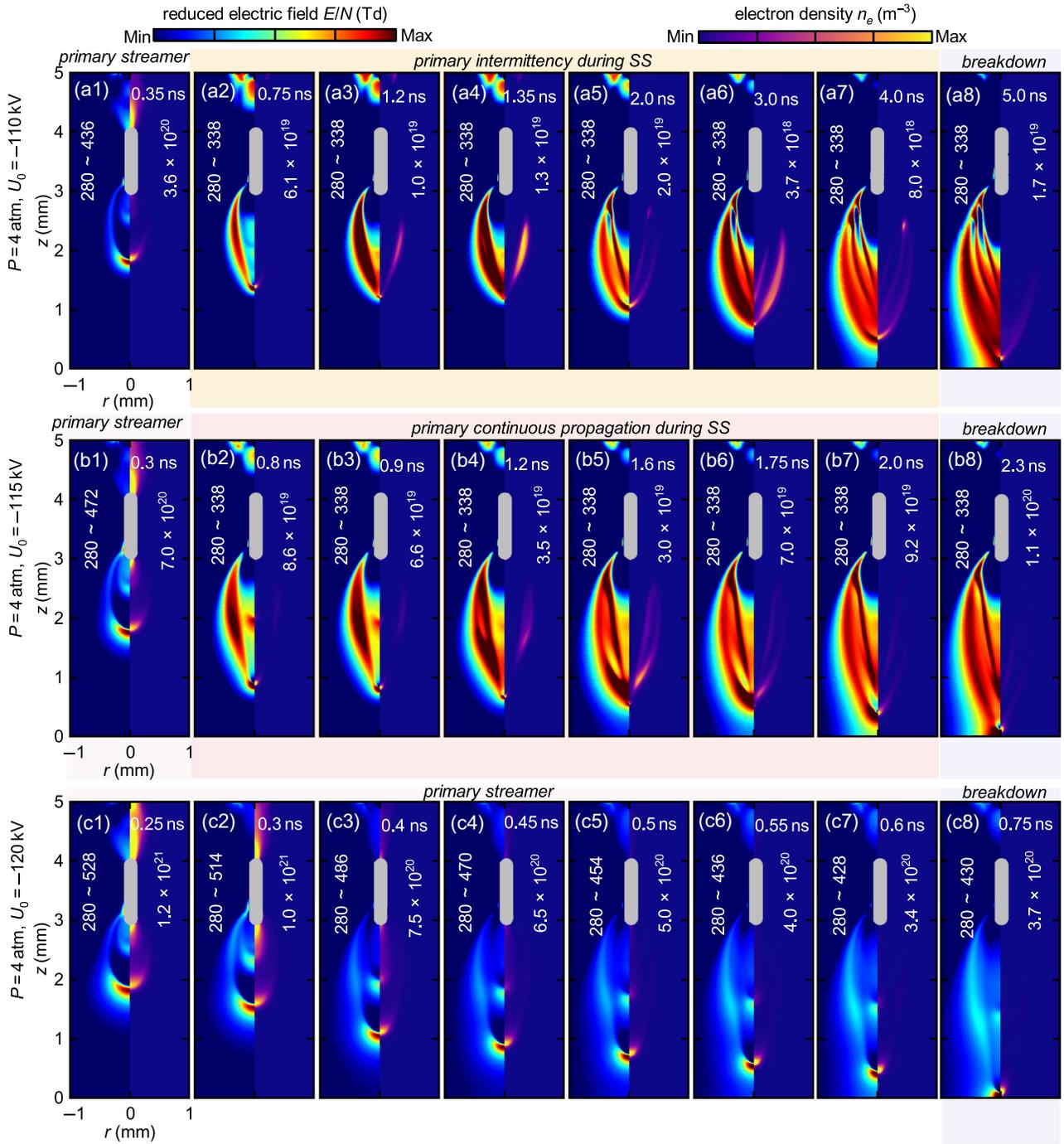

FIG. 11. Evolution of the reduced electric field $E/N$ and electron density $n_e$, illustrating the SS phenomenon under different overvoltage levels: (a1)–(a8) $U_0 = -110$ kV; (b1)–(b8) $U_0 = -115$ kV; and (c1)–(c8) $U_0 = -120$ kV. Herein, $P = 4$ atm, $Q_0 = 0$, and the LFA is utilized with $(E/N)_{cr} = 338$ Td for all configurations. Labels for $(E/N)_{min}$, $(E/N)_{max}$, and $n_{e,max}$ are shown in each panel, where $n_{e,min}$ is fixed at 0.

When the applied voltage is further increased to $U_0 = -120$ kV, corresponding to $(E/N)_b \approx 245$ Td, the primary streamer crosses the gap in just 0.75 ns, as shown in Fig. 11(c8). The rapid propagation significantly reduces the opportunity for the SS mechanism to develop, as shown in Figs. 11(c1)–11(c8). This result indicates that, under extreme overvoltage conditions, characterized by a substantially elevated $(E/N)_b$, the SS mechanism plays a negligible role in the high-pressure $SF_6$ streamer breakdown process.

### E. Preliminary analysis of the particle material

In essence, the influence of different materials of particles on side streamers in $SF_6$ is primarily governed by the electric field distortion factor. Particles with high dielectric constant, e.g., $\varepsilon_r = 4000$, exhibit a field distortion factor that is nearly identical to that of metal particles. Consequently, the applied voltage $U_0$ required to initiate $SF_6$ streamer breakdown is nearly identical for the two cases within the framework of the LFA. Thus, the formation and dynamics of SS induced by high-dielectric-constant particles are essentially the same as those induced by metal particles. In contrast, for particles with low dielectric constant, e.g., $\varepsilon_r = 2.1$, the corresponding field distortion factor is only 22.5% of that of a metal particle. As a result, a significantly higher applied voltage $U_0$ is required to trigger $SF_6$ streamer breakdown, e.g., $U_0 = -160$ kV at 4 atm within the framework of the LFA. The resulting background field $(E/N)_b = 326$ Td significantly exceeds the threshold condition of 245 Td identified in Sec. V D, thereby inhibiting the occurrence of SS.

From a phenomenological perspective, the above analysis indicates a similar manner to the engineering measurements reported by Doepken [85] and Diessner and Trump [9], where insulating particles with low dielectric constant did not alter the insulation strength of $SF_6$. This observation gives rise to the hypothesis that the limited role of insulating particles in $SF_6$ gap breakdown may be attributed to the absence of SS, which otherwise facilitates discharge development. However, a rigorous conclusion requires further validation through both experiments and simulations.

## VI. CONCLUSIONS AND OUTLOOK

### A. Conclusions

In this paper, we simulate and analyze the $SF_6$ streamer breakdown induced by a floating linear metal particle under negative applied voltage. Our key findings are summarized as follows.

(1) *Initial discharge stage.* We investigate the inception of the discharge within the combined gap under the condition $Q_0 = 0$. The discharge inception manifests as a double-end streamer, without a clear sequential initiation between the long and short gaps. This behavior contrasts with floating dielectric ($TiO_2$) discharge or microdischarge theory, where discharge typically initiates in a sequence. Simultaneous ionization at both tips of the metal particle causes electrons from the short gap to conduct to the particle for charging. Subsequently, the redistribution of negative charge due to the particle's electrostatic induction enhances the electric field near the long gap, causing both ends to exceed the streamer threshold.

(2) *Two mechanisms facilitating breakdown.* Two key factors that facilitate breakdown were examined: the role of floating metal particles and the nonlinear breakdown behavior of high-pressure $SF_6$. Based on their microscopic transient behaviors, we propose the following-streamer mechanism [item (3)] and the side-streamer mechanism [item (4)], FS and SS, respectively.

(3) *Role of floating metal particles.* We find that the negative surface charge and surface electric field at the particle's bottom tip continuously increase, similar to the effect of the rising edge of a pulse voltage. This induces two following streamers (FS1 and FS2) in the long gap, with the same polarity as the primary streamer. FS1 is dominated by the charging effect of electron conduction from the short gap and the electrostatic induction of the metal particle; FS1 facilitates breakdown by generating a negative space-charge field, which enhances the electric field at the primary streamer head. FS2 is dominated by the charging effect of positive ions from the long gap and the electrostatic induction of the metal; FS2 facilitates breakdown by merging with FS1, which increases the negative space charge at the FS1 head. This not only accelerates the propagation of FS1 but also further enhances the electric field at the primary streamer head.

(4) *Nonlinear behavior of high-pressure* $SF_6$. We find that the side streamer, a new forward ionization wave, forms along the sides of the primary streamer. The formation of SS is dominated by photoionization, with photoelectrons attaching to the sides and forming negative ions, leading to the enhancement of the side field. SS facilitates breakdown by merging with the primary streamer, thereby increasing the negative space charge at the streamer head and enhancing the head electric field. After merging, if the head field exceeds $(E/N)_{cr}$, the primary streamer exhibits a primary intermittency mode. If the head field remains near but below $(E/N)_{cr}$, the collaboration of photoionization ($S_{ph} > 0$) and effective ionization ($S_{eff} < 0$) drives the forward expansion of the negative space-charge region at the streamer head, leading to a primary interruption mode.

(5) *Pressure dependence of side streamer.* Qualitative analysis indicates that higher pressure results in higher negative-ion production rate $S_n$ at the SS initiation position, making SS more likely to form under high-pressure conditions.

(6) *Effect of overvoltage on side streamer.* Overvoltage provides a higher background field $(E/N)_b$ for the propagation of the primary streamer. When $(E/N)_b \leq 225$ Td, it is insufficient to sustain continuous propagation, and SS merging remains the dominant factor facilitating breakdown. When $(E/N)_b \approx 235$ Td, SS no longer plays a dominant role and acts only supportively. When $(E/N)_b \geq 245$ Td, SS no longer appears, indicating its role to be negligible.

### B. Outlook

It should be acknowledged that the primary limitation of this paper stems from the deterministic 2D fluid model,

which definitely cannot capture certain realistic processes, as discussed in Ref. [86] and references therein. In addition, due to the instabilities in both the following streamers and side streamers—such as the breaking of electric field symmetry during the development of side streamers, and the possible loss of stability in the primary streamer caused by the appearance of following streamers—the discharge morphology will shift into a three-dimensional (3D) form [87,88].

Therefore, we qualitatively compare our results with the impressive experiments investigating the propagation of multiple streamers reported by Nudnova and Starikovskiy [89]. In Ref. [89], multiple streamers occurred more often at higher pressures, consistent with the results of this paper. Ref. [89] also reported four streamer generations within a single voltage pulse, and each new generation propagated using new ways around the space charge of the previous one, which differed from the propagation paths influenced by the merging of streamers in this paper. A possible explanation is that the primary streamer in this paper has not fully crossed the gap, allowing for interaction between the primary streamer and the new streamers. However, a detailed description of the realistic discharge and a rigorous quantitative analysis of streamer stability require fully 3D modeling that accounts for realistic stochastic processes.

## ACKNOWLEDGMENTS

The authors gratefully acknowledge funding support from the National Natural Science Foundation of China (Contract No. 52277154). The authors express their sincere gratitude to the editors and referees for their constructive and insightful comments, which have substantially improved this manuscript. Z.F. thanks Dr. Peng Wang and Caomingzhe Si from the Department of Electrical Engineering, Tsinghua University, for fruitful discussions.

The authors have no conflicts to disclose.

## DATA AVAILABILITY

The data that support the findings of this article are not publicly available. The data are available from the authors upon reasonable request.

## APPENDIX A: GRID SPACING AND TIME STEP

### a. Grid-spacing scheme

For the LMEA, a carefully designed unstructured grid scheme is employed, as shown in Fig. 12, which displays only the discharge region ($0 < r < 1$ mm, $0 < z < 5$ mm), as it is the most important for reproducing the results. Each subregion is highlighted in the figure with specific

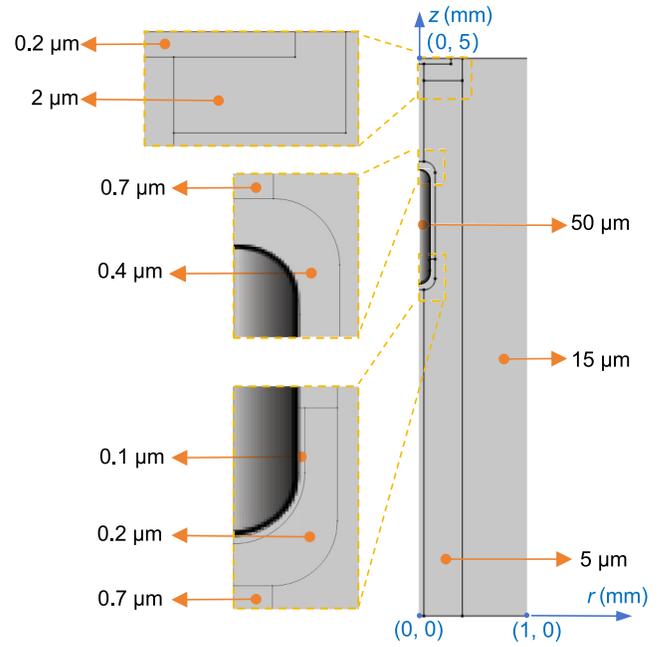

FIG. 12. Grid spacing scheme of the LMEA.

grid spacing. The grids of all regions transition smoothly between each other.

For the LFA, the adaptive mesh refinement is employed to reduce the computational cost. In the entire computational domain, the grid is refined automatically based on

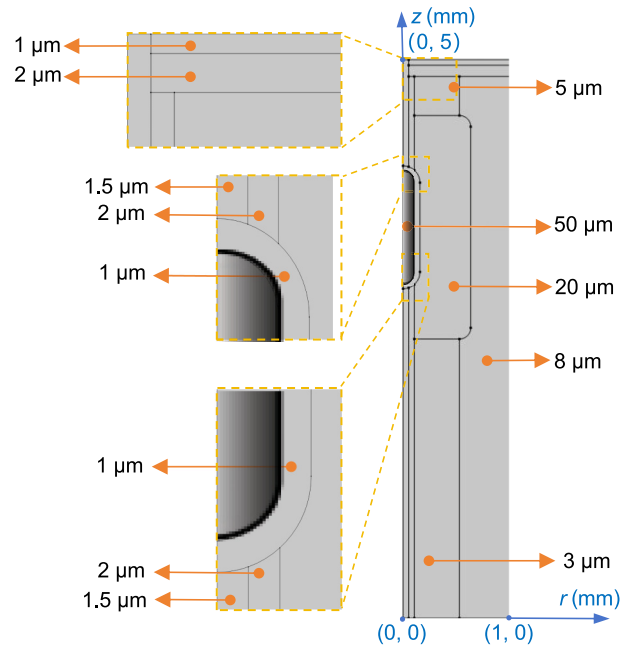

FIG. 13. Initial grid spacing scheme of the LFA.

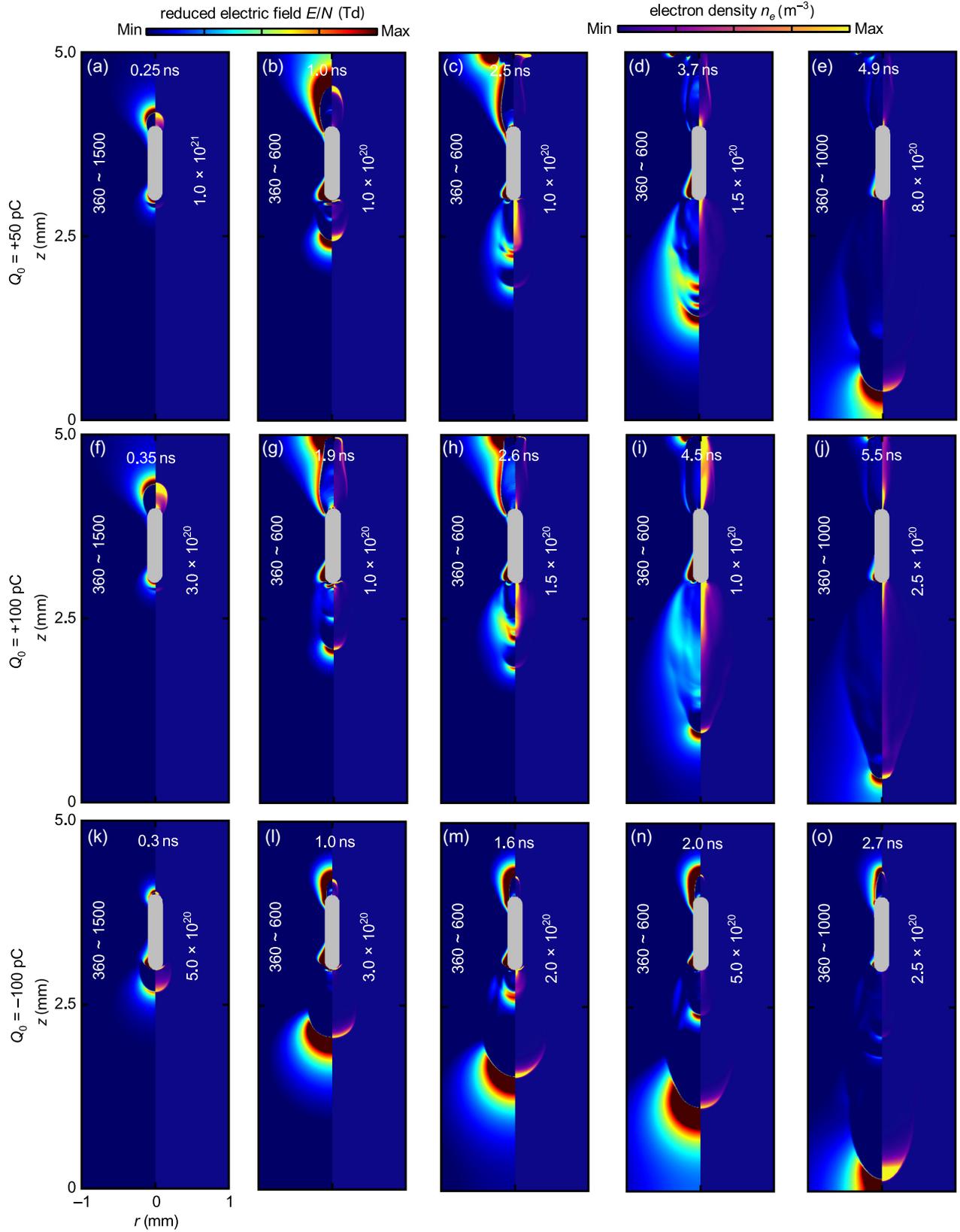

FIG. 14. Evolution of the reduced electric field $E/N$ and electron density $n_e$ for different initial charge: (a)–(e) $Q_0 = +50$ pC; (f)–(j) $Q_0 = +100$ pC; and (k)–(o) $Q_0 = -100$ pC.

the error indicator:

$$\sum_{i \in \{e,p,n\}} \sqrt{\left(\frac{\partial n_i}{\partial r}\right)^2 + \left(\frac{\partial n_i}{\partial z}\right)^2}.$$

The initial grid is shown in Fig. 13, which displays only the discharge region ($0 < r < 1$ mm, $0 < z < 5$ mm). The unstructured grid is employed, with the grids transitioning smoothly between all regions. During the refinement, the minimum grid spacing is 0.1 µm. The simulations are performed using two Intel® Xeon® Gold 6246R @ 3.4 GHz and 12 memory channels, each equipped with 64 GB of DDR4-2933 RAM.

*b. Time-step scheme*

The implicit backward differentiation formula (BDF) method is used, with a maximum BDF order of 2 and a minimum BDF order of 1. The relative tolerance is set to $10^{-2}$. A parallel sparse direct solver (PARDISO) is chosen as the direct linear system solver. Notably, the maximum time step is limited to 0.5 ps; the rationale will be presented elsewhere. For insights into the influence of time step, one can refer to the impressive work reported by Bagheri *et al.* [44].

### APPENDIX B: PRELIMINARY STUDY OF THE EFFECT OF INITIAL CHARGE $Q_0$ ON FOLLOWING STREAMERS

In the cases of positive charge $Q_0 = +50$ and $+100$ pC, the Laplacian field near the bottom tip of the metal particle weakens. This weakening slows the propagation of the primary streamer across the gap. Consequently, the breakdown time increases significantly, from 3.2 ns for $Q_0 = 0$ to 4.9 and 5.5 ns for $Q_0 = +50$ and $+100$ pC, respectively. Additionally, the facilitation effect of following streamers on the breakdown process becomes more pronounced. Specifically, as the primary streamer propagates away from the particle's tip, the weakening background field causes its head field to drop below 600 Td [see Figs. 14(c) and 14(h)]. Under the facilitation effect of FS, the head field of the primary streamer subsequently increases to ~850 Td [see Figs. 14(d) and 14(i)]. This field recovery highlights the critical role of FS in facilitating combined-gap breakdown. In addition, more FS events occur in these two cases. After the merging of FS1 and FS2 [see Figs. 14(c) and 14(h)], a new following streamer FS3 forms [see Figs. 14(d) and 14(i)], further advancing the primary streamer.

In the cases of negative charge $Q_0 = -100$ pC, the enhanced Laplacian electric field near the bottom tip accelerates the propagation of the primary streamer in the long gap. Combined with the influence of FS, this leads to breakdown occurring in the long gap at 2.7 ns, before the short gap breaks down. However, this outcome reflects a limitation of this paper rather than a real-world operating scenario. Specifically, the simplified geometry used significantly shortens the long gap compared to that in the real-world scenario, as discussed in Sec. II A. For more realistic analyses, future work should adopt a larger gap ratio ($g_l/g_s$).

[1] J. Wang, Q. Hu, Y. Chang, J. Wang, R. Liang, Y. Tu, C. Li, and Q. Li, Metal particle contamination in gas-insulated switchgears/gas-insulated transmission lines, CSEE J. Power Energy Syst. **7**, 1011 (2021).

[2] M. Seeger, F. Macedo, U. Riechert, M. Bujotzek, A. Hassanpoor, and J. Häfner, Trends in high voltage switchgear research and technology, IEEJ Trans. Electr. Electron. Eng. **20**, 322 (2025).

[3] C. Li *et al.*, China's 10-year progress in DC gas-insulated equipment: From basic research to industry perspective, iEnergy **1**, 400 (2022).

[4] W. Zhuang, Z. Liang, Y. Yi, W. Qin, F. Liang, X. Fan, T. Ma, J. Hu, C. Li, B. Zhang, and J. He, Metallic particles in DC gas-insulated transmission lines, J. Phys. D: Appl. Phys. **58**, 233001 (2025).

[5] J. R. Laghari and A. H. Qureshi, A review of particle-contaminated gas breakdown, IEEE Trans. Electr. Insul. **EI-16**, 388 (1981).

[6] H. Kuwahara, S. Inamura, T. Watanabe, and Y. Arahata, Effect of solid impurities on breakdown in compressed $SF_6$ gas, IEEE Trans. Power App. Syst. **PAS-93**, 1546 (1974).

[7] L. Donzel, M. Seeger, D. Over, and J. Carstensen, Metallic particle motion and breakdown at AC voltages in $CO_2/O_2$ and $SF_6$, Energies **15**, 2804 (2022).

[8] S. Xiao, X. Zhang, R. Zhuo, D. Wang, J. Tang, S. Tian, and Y. Li, The influence of Cu, Al and Fe free metal particles on the insulating performance of $SF_6$ in C-GIS, IEEE Trans. Dielectr. Electr. Insul. **24**, 2299 (2017).

[9] A. Diessner and J. G. Trump, Free conducting particles in a coaxial compressed-gas-insulated system, IEEE Trans. Power App. Syst. **PAS-89**, 1970 (1970).

[10] H. You, Q. Zhang, C. Guo, P. Xu, J. Ma, Y. Qin, T. Wen, and Y. Li, Motion and discharge characteristics of metal particles existing in GIS under DC voltage, IEEE Trans. Dielectr. Electr. Insul. **24**, 876 (2017).

[11] J. Ma, Q. Zhang, Z. Wu, C. Guo, T. Wen, G. Wang, and C. Gao, Breakdown characteristics of particle-contaminated HVDC GIL under superimposed voltage of DC and impulse, IEEE Trans. Dielectr. Electr. Insul. **25**, 1439 (2018).

[12] J. Wang, Q. Li, B. Li, C. Chen, S. Liu, and C. Li, Theoretical and experimental studies of air gap breakdown triggered by free spherical conducting particles in DC uniform field, IEEE Trans. Dielectr. Electr. Insul. **23**, 1951 (2016).

[13] X. Li, X. Hu, H. Xu, P. Jiang, M. Wu, B. Niu, Z. Li, and Q. Zhang, "Multi-particle-in-series" phenomenon and discharge properties induced by multiple free particles in GIS, IEEE Trans. Power Deliv. **38**, 4039 (2023).

[14] W. Zhong, Y. Shi, C. Zhang, and X. Li, Prediction of microparticle-initiated breakdown in mm-scaled air gap

based on Townsend theory and streamer inception, IEEE Trans. Dielectr. Electr. Insul. **27**, 1095 (2020).

[15] Y. Shi, W. Zhong, A. Xu, and P. Gan, Study on the mechanism of microparticle-initiated breakdown in the air gap, IEEE Trans. Plasma Sci. **49**, 813 (2021).

[16] Q. Sun, Q. Zhou, W. Yang, Y. Dong, H. Zhang, M. Song, and Y. Wu, Theoretical and numerical studies of breakdown phenomena triggered by microparticle in nitrogen gaps, Plasma Sources Sci. Technol. **30**, 045001 (2021).

[17] N. Lebedev and I. Skalskaya, Force acting on a conducting sphere in field of a parallel plate condenser, Sov. Phys. Tech. Phys. **7**, 268 (1962).

[18] Z. Feng, Y. Jiang, L. Zhang, Z. Liu, K. Wang, X. Wang, X. Zou, H. Luo, and Y. Fu, Microscopic characteristics of $SF_6$ partial discharge induced by a floating linear metal particle, Appl. Phys. Lett. **125**, 134101 (2024).

[19] R. T. Waters, Electrical breakdown at high pressures: A Paschen law function and compressible gas dynamics, J. Phys. D: Appl. Phys. **52**, 025203 (2018).

[20] M. Seeger and M. Clemen, Partial discharges and breakdown in $SF_6$ in the pressure range 25–150 kPa in nonuniform background fields, J. Phys. D: Appl. Phys. **47**, 025202 (2013).

[21] I. D. Chalmers, I. Gallimberti, A. Gibert, and O. Farish, The development of electrical leader discharges in a point-plane gap in $SF_6$, Proc. R. Soc. Lond. A **412**, 285 (1987).

[22] M. Seeger, L. Niemeyer, and M. Bujotzek, Leader propagation in uniform background fields in $SF_6$, J. Phys. D: Appl. Phys. **42**, 185205 (2009).

[23] A. H. Cookson, Review of high-voltage gas breakdown and insulators in compressed gas, IEE Proc. A Sci. Meas. Technol. **128**, 303 (1981).

[24] A. H. Cookson, O. Farish, and G. M. L. Sommerman, Effect of conducting particles on AC corona and breakdown in compressed $SF_6$, IEEE Trans. Power App. Syst. **PAS-91**, 1329 (1972).

[25] A. H. Cookson and O. Farish, Particle-initiated breakdown between coaxial electrodes in compressed $SF_6$, IEEE Trans. Power App. Syst. **PAS-92**, 871 (1973).

[26] C. Cooke, R. Wootton, and A. Cookson, Influence of particles on AC and DC electrical performance of gas insulated systems at extra-high-voltage, IEEE Trans. Power App. Syst. **96**, 768 (1977).

[27] I. Gallimberti and N. Wiegart, Streamer and leader formation in $SF_6$ and $SF_6$ mixtures under positive impulse conditions. II. Streamer to leader transition, J. Phys. D: Appl. Phys. **19**, 2363 (1986).

[28] Z. Zhao, Z. Dai, A. Sun, and J. Li, Streamer-to-precursor transition in $N_2$–$SF_6$ mixtures under positive repetitive submicrosecond pulses, High Volt. **7**, 382 (2022).

[29] Z. Zhao, Z. Huang, X. Zheng, C. Li, A. Sun, and J. Li, Evolutions of repetitively pulsed positive streamer discharge in electronegative gas mixtures at high pressure, Plasma Sources Sci. Technol. **31**, 075006 (2022).

[30] Z. Wu, Q. Zhang, L. Zhang, C. Guo, Q. Du, and L. Pang, Neglected culprit of nonlinear discharge characteristics in $SF_6$: Shielding effect induced by positive glow corona discharge, Plasma Sources Sci. Technol. **28**, 085018 (2019).

[31] X. Meng, W. Zhuang, L. Lin, H. Li, Z. Cao, and H. Mei, Characteristics of streamer discharge in $SF_6$ gas under different pressures, IEEE Trans. Dielectr. Electr. Insul., 1 (2024).

[32] Q. Gao, C. Niu, X. Wang, A. Yang, Y. Wu, A. B. Murphy, M. Rong, X. Fu, J. Liu, and Y. Xu, Chemical kinetic modeling and experimental study of $SF_6$ decomposition byproducts in 50 Hz ac point-plane corona discharges, J. Phys. D: Appl. Phys. **51**, 295202 (2018).

[33] L. Zhang, Z. Liu, Y. Guo, J. Liu, K. Wang, H. Luo, and Y. Fu, Kinetic model of grating-like DBD fed with flowing humid air, Plasma Sources Sci. Technol. **33**, 025001 (2024).

[34] D. Levko and Y. E. Krasik, Numerical simulation of runaway electrons generation in sulfur hexafluoride, J. Appl. Phys. **111**, 013305 (2012).

[35] D. Levko, V. Tz. Gurovich, and Y. E. Krasik, Conductivity of nanosecond discharges in nitrogen and sulfur hexafluoride studied by particle-in-cell simulations, J. Appl. Phys. **111**, 123303 (2012).

[36] Z. Xiong and M. J. Kushner, Branching and path-deviation of positive streamers resulting from statistical photon transport, Plasma Sources Sci. Technol. **23**, 065041 (2014).

[37] Y. Guo and S. Nijdam, Statistical analysis on branching characteristics of positive streamer discharges in $N_2$–$O_2$ mixtures, Plasma Sources Sci. Technol. **33**, 045006 (2024).

[38] Z. Wang, S. Dijcks, Y. Guo, M. van der Leegte, A. Sun, U. Ebert, S. Nijdam, and J. Teunissen, Quantitative modeling of streamer discharge branching in air, Plasma Sources Sci. Technol. **32**, 085007 (2023).

[39] X. Li, S. Dijcks, A. Sun, S. Nijdam, and J. Teunissen, Investigation of positive streamers in $CO_2$: Experiments and 3D particle-in-cell simulations, Plasma Sources Sci. Technol. **33**, 095009 (2024).

[40] Z. Wang, A. Sun, S. Dujko, U. Ebert, and J. Teunissen, 3D simulations of positive streamers in air in a strong external magnetic field, Plasma Sources Sci. Technol. **33**, 025007 (2024).

[41] B. Guo, U. Ebert, and J. Teunissen, 3D particle-in-cell simulations of negative and positive streamers in $C_4F_7N$–$CO_2$ mixtures, Plasma Sources Sci. Technol. **32**, 115001 (2023).

[42] R. Marskar, Stochastic and self-consistent 3D modeling of streamer discharge trees with kinetic Monte Carlo, J. Comput. Phys. **504**, 112858 (2024).

[43] R. Marskar, A 3D kinetic Monte Carlo study of streamer discharges in $CO_2$, Plasma Sources Sci. Technol. **33**, 025023 (2024).

[44] B. Bagheri, J. Teunissen, U. Ebert, M. M. Becker, S. Chen, O. Ducasse, O. Eichwald, D. Loffhagen, A. Luque, D. Mihailova, J. M. Plewa, J. van Dijk, and M. Yousfi, Comparison of six simulation codes for positive streamers in air, Plasma Sources Sci. Technol. **27**, 095002 (2018).

[45] Y. Zhu, X. Chen, Y. Wu, J. Hao, X. Ma, P. Lu, and P. Tardiveau, Simulation of ionization-wave discharges: A direct comparison between the fluid model and E-FISH measurements, Plasma Sources Sci. Technol. **30**, 075025 (2021).

[46] P. Viegas, E. Slikboer, Z. Bonaventura, O. Guaitella, A. Sobota, and A. Bourdon, Physics of plasma jets and interaction with surfaces: review on modelling and experiments, Plasma Sources Sci. Technol. **31**, 053001 (2022).

[47] N. Y. Babaeva and M. J. Kushner, Effect of inhomogeneities on streamer propagation: I. Intersection with

ZIHAO FENG *et al.*

[47] isolated bubbles and particles, Plasma Sources Sci. Technol. **18**, 035009 (2009).

[48] N. Y. Babaeva and M. J. Kushner, Effect of inhomogeneities on streamer propagation: II. Streamer dynamics in high pressure humid air with bubbles, Plasma Sources Sci. Technol. **18**, 035010 (2009).

[49] L. G. Christophorou and J. K. Olthoff, Electron interactions with $SF_6$, J. Phys. Chem. Ref. Data **29**, 267 (2000).

[50] G. J. M. Hagelaar and L. C. Pitchford, Solving the Boltzmann equation to obtain electron transport coefficients and rate coefficients for fluid models, Plasma Sources Sci. Technol. **14**, 722 (2005).

[51] R. Van Brunt and J. Herron, Fundamental processes of $SF_6$ decomposition and oxidation in glow and corona discharges, IEEE Trans. Electr. Insul. **25**, 75 (1990).

[52] F. Zeng, M. Zhang, D. Yang, and J. Tang, Hybrid numerical simulation of decomposition of $SF_6$ under negative DC partial discharge process, Plasma Chem. Plasma Process. **39**, 205 (2019).

[53] D. Levko and L. L. Raja, Fluid versus global model approach for the modeling of active species production by streamer discharge, Plasma Sources Sci. Technol. **26**, 035003 (2017).

[54] D. Levko and L. L. Raja, Computational analysis of electrical breakdown of $SF_6/N_2$ mixtures, J. Appl. Phys. **133**, 053301 (2023).

[55] R. Morrow, A survey of the electron and ion transport properties of $SF_6$, IEEE Trans. Plasma Sci. **14**, 234 (1986).

[56] M. B. Zhelezniak, A. K. Mnatsakanian, and S. V. E. Sizykh, Photoionization of nitrogen and oxygen mixtures by radiation from a gas discharge, High Temp. Sci. **20**, 357 (1982).

[57] A. Bourdon, V. P. Pasko, N. Y. Liu, S. Célestin, P. Ségur, and E. Marode, Efficient models for photoionization produced by non-thermal gas discharges in air based on radiative transfer and the Helmholtz equations, Plasma Sources Sci. Technol. **16**, 656 (2007).

[58] A. Luque, U. Ebert, C. Montijn, and W. Hundsdorfer, Photoionization in negative streamers: Fast computations and two propagation modes, Appl. Phys. Lett. **90**, 081501 (2007).

[59] M. S. Bhalla and J. D. Craggs, Measurement of ionization and attachment coefficients in sulphur hexafluoride in uniform fields, Proc. Phys. Soc. **80**, 151 (1962).

[60] L. E. Kline, D. K. Davies, C. L. Chen, and P. J. Chantry, Dielectric properties for $SF_6$ and $SF_6$ mixtures predicted from basic data, J. Appl. Phys. **50**, 6789 (1979).

[61] H. Boyd and G. Crichton, Measurement of ionisation and attachment coefficients in $SF_6$, Proc. Inst. Electr. Eng. **118**, 1872 (1971).

[62] V. N. Maller and M. S. Naidu, Ratio of diffusion coefficient to mobility for electrons in $SF_6$-air and freon-nitrogen mixtures, IEEE Trans. Plasma Sci. **3**, 205 (1975).

[63] T. Yoshizawa, Y. Sakai, H. Tagashira, and S. Sakamoto, Boltzmann equation analysis of the electron swarm development in $SF_6$, J. Phys. D: Appl. Phys. **12**, 1839 (1979).

[64] J. P. Novak and M. F. Fréchette, Transport coefficients of $SF_6$ and $SF_6$–$N_2$ mixtures from revised data, J. Appl. Phys. **55**, 107 (1984).

[65] M. C. Siddagangappa, C. S. Lakshminarasimha, and M. S. Naidu, Electron attachment in binary mixtures of electronegative and buffer gases, J. Phys. D: Appl. Phys. **15**, L83 (1982).

[66] K. B. McAfee Jr and D. Edelson, Identification and mobility of ions in a Townsend discharge by time-resolved mass spectrometry, Proc. Phys. Soc. **81**, 382 (1963).

[67] S. A. Madhar, P. Mraz, A. R. Mor, and R. Ross, Physical interpretation of the floating electrode defect patterns under AC and DC stress conditions, Int. J. Electr. Power Energy Syst. **118**, 105733 (2020).

[68] P. Wenger, M. Beltle, S. Tenbohlen, U. Riechert, and G. Behrmann, Combined characterization of free-moving particles in HVDC-GIS using UHF PD, high-speed imaging, and pulse-sequence analysis, IEEE Trans. Power Deliv. **34**, 1540 (2019).

[69] D. Fahmi, H. A. Illias, H. Mokhlis, and I. M. Y. Negara, Particle-triggered corona discharge characteristics in air insulation under DC voltage, IEEE Trans. Dielectr. Electr. Insul. **30**, 658 (2023).

[70] H. Ji, C. Li, Z. Pang, G. MA, X. Cui, W. Zhao, and J. Wang, Influence of tip corona of free particle on PD patterns in GIS, IEEE Trans. Dielectr. Electr. Insul. **24**, 259 (2017).

[71] H. Ji, C. Li, Z. Pang, G. Ma, X. Cui, Z. Zeng, and Z. Rong, Moving behaviors and harmfulness analysis of multiple linear metal particles in GIS, IEEE Trans. Dielectr. Electr. Insul. **23**, 3355 (2016).

[72] Y. Wang, R. L. Champion, L. D. Doverspike, J. K. Olthoff, and R. J. Van Brunt, Collisional electron detachment and decomposition cross sections for $SF_6^-$, $SF_6^-$, and $F^-$ on $SF_6$ and rare gas targets, J. Chem. Phys. **91**, 2254 (1989).

[73] R. J. Van Brunt and M. Misakian, Mechanisms for inception of DC and 60-Hz AC corona in $SF_6$, IEEE Trans. Electr. Insul. **EI-17**, 106 (1982).

[74] M. Seeger, L. Niemeyer, and M. Bujotzek, Partial discharges and breakdown at protrusions in uniform background fields in $SF_6$, J. Phys. D: Appl. Phys. **41**, 185204 (2008).

[75] S. Mirpour and S. Nijdam, Experimental investigation on streamer inception from artificial hydrometeors, Plasma Sources Sci. Technol. **31**, 105009 (2022).

[76] A. Abahazem, N. Merbahi, O. Ducasse, O. Eichwald, and M. Yousfi, Primary and secondary streamer dynamics in pulsed positive corona discharges, IEEE Trans. Plasma Sci. **36**, 924 (2008).

[77] R. Ono and T. Oda, Formation and structure of primary and secondary streamers in positive pulsed corona discharge—effect of oxygen concentration and applied voltage, J. Phys. D: Appl. Phys. **36**, 1952 (2003).

[78] Y. Li, E. M. van Veldhuizen, G. J. Zhang, U. Ebert, and S. Nijdam, Positive double-pulse streamers: How pulse-to-pulse delay influences initiation and propagation of subsequent discharges, Plasma Sources Sci. Technol. **27**, 125003 (2018).

[79] A. Komuro, R. Ono, and T. Oda, Effects of pulse voltage rise rate on velocity, diameter and radical production of an atmospheric-pressure streamer discharge, Plasma Sources Sci. Technol. **22**, 045002 (2013).

[80] A. Luque, V. Ratushnaya, and U. Ebert, Positive and negative streamers in ambient air: Modelling evolution and velocities, J. Phys. D: Appl. Phys. **41**, 234005 (2008).

[81] Z. Zhao, Z. Wang, Z. Duan, and Y. Fu, Dynamical similarity of streamer propagation in geometrically similar combined air gaps, IEEE Trans. Dielectr. Electr. Insul. **32**, 1017 (2025).

[82] F. Wang, K. Liang, L. Zhong, S. Chen, T. Wan, X. Duan, and X. Dong, Effect of metal ions and suspended particles on streamer propagation, IEEE Trans. Dielectr. Electr. Insul. **30**, 1154 (2023).

[83] S. Wang, Y. Xu, and F. Lü, Simulation study of the effect of microscopic disturbances on the propagation of initial streamer and branching characteristics in transformer oil, IEEE Trans. Dielectr. Electr. Insul. **31**, 75 (2024).

[84] Z. Wu, B. Lin, X. Fan, Q. Zhang, and L. Li, Electric field dependence of $SF_6$ nonlinear discharge characteristics: $N$-curve estimations, Plasma Sources Sci. Technol. **30**, 015009 (2021).

[85] H. C. Doepken, Compressed-gas insulation in large coaxial systems, IEEE Trans. Power App. Syst. **PAS-88**, 364 (1969).

[86] A. Y. Starikovskiy, N. L. Aleksandrov, and M. N. Shneider, Simulation of decelerating streamers in inhomogeneous atmosphere with implications for runaway electron generation, J. Appl. Phys. **129**, 063301 (2021).

[87] A. Y. Starikovskiy and N. L. Aleksandrov, "Gas-dynamic diode": Streamer interaction with sharp density gradients, Plasma Sources Sci. Technol. **28**, 095022 (2019).

[88] A. Y. Starikovskiy and N. L. Aleksandrov, Blocking streamer development by plane gaseous layers of various densities, Plasma Sources Sci. Technol. **29**, 034002 (2020).

[89] M. M. Nudnova and A. Y. Starikovskii, Development of streamer flash initiated by HV pulse with nanosecond rise time, IEEE Trans. Plasma Sci. **36**, 896 (2008).